# Exploring associations between micro-level models of innovation diffusion and emerging macro-level adoption patterns

Carlos E. Laciana[a], Santiago L. Rovere[a] and

Guillermo P. Podestá[b]

[a] Grupo de Aplicaciones de Modelos de Agentes (GAMA),
Facultad de Ingeniería, Universidad de Buenos Aires
Avenida Las Heras 2214, Ciudad Autónoma de Buenos Aires
C1127AAR, Argentina.
clacian@fi.uba.ar

[b] Rosenstiel School of Marine and Atmospheric Science,
University of Miami, 4600 Rickenbacker Causeway, Miami, FL
33149-1098, USA.

September 21, 2012

# Abstract

A micro-level agent-based model of innovation diffusion was developed that explicitly combines (a) an individual's perception of the advantages or relative utility derived from adoption, and (b) social influence from members of the individual's social network. The micro-model was used to simulate macro-level diffusion patterns emerging from different configurations of micro-model parameters. Micro-level simulation results matched very closely the adoption patterns predicted by the widely-used Bass macro-level model (Bass, 1969 [1]). For a portion of the $p-q$ domain, results from micro-simulations were consistent with aggregate-level adoption patterns reported in the literature. Induced Bass macro-level parameters $p$ and $q$ responded to changes in micro-parameters: (1) $p$ increased with the number of innovators and with the rate at which innovators are introduced; (2) $q$ increased with the probability of rewiring in small-world networks, as the characteristic path length decreases; and (3) an increase in the overall perceived utility of an innovation caused a corresponding increase in induced $p$ and $q$ values. Understanding micro to macro linkages can inform the design and assessment of marketing interventions on micro-variables – or processes related to them – to enhance adoption of future products or technologies.

# 1  Introduction

The rapid pace of technological innovation and its importance to the global economy has focused the attention of academia and industry on understanding the processes and drivers behind the diffusion and adoption of innovations [2]. Innovations are defined here as a new product, a new process, a new technology, or even a new organizational form. In turn, the spread of an innovation in a market is termed "diffusion." From the marketing perspective, it is important to understand how marketing communication efforts and social influence from previous adopters may affect the adoption decisions of consumers and, consequently, the diffusion of a new product. The wealth of research into modeling and forecasting the diffusion of innovations is impressive; recent reviews of diffusion models can be found in [3-5].

The diffusion of innovations may be approached from two alternative perspectives: the macroscopic and microscopic. At the macro level, an entire market is examined to identify or forecast how many customers will eventually adopt an innovation (the market size), and when they will adopt (the time path of adoption). Many macro-level studies of innovation diffusion are rooted on the influential work of Bass [1] described in more detail in the next section. Macroscopic models provide parsimonious and analytically tractable ways to look at a whole market and interpret its behavior. A related advantage is their use of market-level data – often more available than individual-level data – to forecast sales [6]. On the other hand, macro-models do not provide insight about the processes that determine adoption, or on how individual market interactions are linked to global market behavior [5].

In contrast, at the microscopic level each decision unit (an individual, a household, a firm) must choose whether to adopt an innovation; in this approach, analytical emphasis is placed on understanding the processes and factors influencing the individual adoption behavior, including both product characteristics and social interactions and to analyzing how it affects the aggregate diffusion process [7, 8]; understanding the nature of these processes can inform marketing strategy recommendations [9]

In the last few decades there has been growing awareness of the importance of social structure as the substrate for the diffusion of innovations. An implicit assumption in macro approaches such as the Bass model is that the target population is fully-connected, that is, that every individual potentially can interact with everyone else in the population and can exert the same social influence as everyone else [10]. This is clearly not realistic, as there is considerable evidence that social networks are neither homogeneous nor fully connected. In particular, the topologies known as small world networks (SWNs, [11]) appear often in models of social relations [12]. SWNs have high values of both connectivity (i.e., short average path length) and clustering, making propagation of information more efficient than in other topologies [13, 14].

In addition to social structure, recent strands of the diffusion literature also emphasize heterogeneity in the characteristics of consumers – such as their susceptibility to the behavior of others or sensitivity to price – that lead to differences in an individual's propensity to adopt [8]. Recent studies have even challenged the prevailing notion that social contagion is an important driver of new product diffusion, instead pointing out that typical S-shaped diffusion curves need not stem from social contagion, but can result from heterogeneity among individuals in their intrinsic tendency to adopt [9]. Consumer heterogeneity, however, is not explicitly considered in macro-level diffusion models.

Simulation models (e.g., cellular automata, agent-based models, percolation models) provide a way to systematically conduct experiments on how micro- level variables affect innovation diffusion processes [8]. Recently, agent-based modeling [15, 16] has increasingly been used in diffusion studies because it can overcome some of the limitations of aggregate-level models such as the assumption of homogeneous adopters, or the lack of explicit social structure. Reviews of agent-based modeling in the context of innovation diffusion are in [17] and [6].

In agent-based models (ABMs) of innovation diffusion the modeling unit is the individual consumer or agent, not the social system as a whole. The micro-level processes that drive adoption decisions are explicitly specified. In turn, macro-level adoption dynamics emerge from the aggregated individual behavior and the interactions between agents [18]. Moreover, ABMs can capture individual heterogeneity in several characteristics, including responsiveness to price and advertising [19], presence of negative word of mouth [20], intrinsic consumer innovativeness [21], and individual roles in the social network – that is, hubs, connectors, and experts [22]. In ABMs, agents can interact with other agents through social networks that can be explicitly specified with different topologies and parameters. The ABM approach thus allows definition of a broader range of social interactions than Bass' "word of mouth". For instance, [23] expanded adoption decision rules in order to reflect network externalities that exist when consumers derive utility from a product based on the number of other users.

The central goal of this paper is to explore associations between parameters of a micro-level ABM and emergent patterns [18] from widely-used, macro-level models of innovation diffusion. Previous studies linking individual-level behavior and market-level patterns have been undertaken by [9, 24, 25]. In particular, the relationship between ABMs and the Bass model was studied by [26] and [7]. Shaikh et al. [27] showed that adoption by agents connected by a small-world network can be aggregated to create the Bass model. However, the interface between the individual level and the aggregate level still needs further exploration. We show here that results from an ABM can be consistent with the aggregate-level empirical data about adoption that are typically more available for analysis [23]. From a theoretical point of view, our contribution is a micro-level approach that considers plausible social network topologies, and allows heterogeneity among decision-makers (not explored here for the sake of length). Moreover, the decision algorithm underlying our approach [28] would let us easily introduce uncertainty in adoption decisions (e.g., due to social and economic contexts). The combination of all these features offers a versatile tool for future work.

The paper is organized as follows: section 2 provides a brief description of the Bass model and its associated parameters. Section 3 introduces the micro-level agent-based model we developed, including details about the adoption algorithm and the implementation that allowed numerical experiments. Section 4 describes experiments performed to link the micro and macro levels. We study how changes in the topology of the social network can affect the induced parameters of the Bass model and we estimate analytically how takeoff time changes. This allows us to understand the influence of micro-level processes on patterns of adoption at the macro level. Possible application of this understanding is illustrated with an example. Section 5 summarizes the main conclusions, and points to possible future work.

## 2  Brief overview of the Bass model

A large body of research using macroscopic models of innovation diffusion has been based on the framework originally developed by Bass [1]. The Bass model characterizes the diffusion of a product or technology as a contagious process initiated by the spontaneous adoption of consumers responding to external influences (such as mass media coverage), and propelled by internal influences (such as word-of-mouth between individuals). Although originally developed for consumer durable goods, the Bass model has described successfully the diffusion dynamics of a wide spectrum of products [1, 29-33].

The Bass model involves two main parameters: the innovation parameter $p$ that reflects people's intrinsic tendency to adopt an innovation, and the imitation parameter $q$ that reflects "word of mouth" or "social contagion." These two parameters respectively capture the behaviors of two types of adopters in a population: innovators and imitators. Innovators have a constant propensity to adopt; they accept a new product or technology because they perceive it to have comparative advantages or a positive difference in utility against existing alternatives. The innovators' decision to adopt is mostly influenced by external influences (e.g., advertising). In contrast, the adoption decision of imitators is determined by internal influences resulting from interactions among adopters and potential adopters in the social system; specifically, adoption by imitators is influenced by the proportion of previous adopters.

In the original Bass article [1] and many of the studies that followed it, $q$ was interpreted as representing the influence of word of mouth among individuals. This word of mouth – now facilitated by multiple technologies and media (e.g., e-mail, Twitter, YouTube) – includes both functional and social signals [10]. Functional signals convey market perceptions of the functional attributes of a product, such as its quality or the risks involved in adoption. Correspondingly, social signals also transmitted by word of mouth convey norms concerning consumptive behavior and information on the social consequences of adopting the product, including the social risks of adopting (or not) the innovation [34].

The temporal pattern of adoption is parsimoniously represented in the Bass model by a differential equation (1) that can be solved analytically [1, 35]:

$$dn^+ / dt = (p + q n^+)(1 - n^+) \ . \tag{1}$$

At any given time, all individuals are assumed to be in one of two possible states. An individual is in state + if s/he already has adopted the innovation. Conversely, state - denotes a non-adopter. In Eq. (1), $n^+$ is the proportion of the target population in state + (i.e., the total number of adopters divided by the population size). In all calculations, we assume that at the start of the simulations all individuals are in state -, that is, $n^+(0) = 0$. Solving Eq. (1) assuming this initial condition, we obtain:

$$n^+(t) = \frac{1 - e^{-(p+q)t}}{1 + (q/p)\, e^{-(p+q)t}} \ . \tag{2}$$

Parameters $p$ and $q$ can be estimated from actual adoption data. For $p$ and $q$ values reported in the literature, the proportion of adopters as a function of time generally follows an S-shaped function.

An important observable quantity widely discussed in the literature is the "takeoff time" $t_{TO}$, defined as the time in the life-cycle of an innovation when it transitions from the introduction stage to the growth stage [36]. In other words, the takeoff time represents the change from a "desirable product" to a "popular product." Mathematically, $t_{TO}$ corresponds to the time when the change in the derivative of the proportion of adopters reaches a maximum [36]. Therefore, $t_{TO}$ can be calculated solving the equation:

$$d^3 n^+ / dt^3 = 0 \ . \tag{3}$$

The solution to Eq. (3), which had already been derived by [37] is

$$t_{TO} = \frac{1}{p+q} \ln\left(\frac{q}{(2+\sqrt{3})}\right) \ . \tag{4}$$

The takeoff time is very important to a firm introducing an innovation, as a fast and substantial takeoff can guarantee a competitive advantage, set up a wave of contagious consumption, and therefore determine whether an innovation becomes a hit or a flop. On the other hand, rapid growth requires extensive resources that need to be available, such as sales staff, and manufacturing, distribution, and inventory support [38, 39]. Most importantly, takeoff is often a signal of the mass adoption of a product and its ultimate commercial success. Knowing the impact of company decisions on the likelihood and timing of takeoff is important for effectively managing such success [40].

## 3 Agent-based modeling of innovation diffusion

To explore the connection between micro- and macro-level modeling approaches to the diffusion of innovations we implement an ABM of innovation adoption based on the well-known Ising model [28, 41]. The model explicitly combines (a) an individual's perception of the advantages or relative utility derived from adoption of an innovation, and (b) social influence from relevant members of the individual's social network. We embed these micro-level dynamics into an ABM to simulate emerging macro-level patterns of innovation diffusion throughout aggregate systems with different configurations (temporal and spatial distributions of innovators, alternative topologies of the social network).

## 3.1 Innovation adoption algorithm

The adoption algorithm in our ABM has been described in detail elsewhere [28] and will only be briefly reviewed here. At each time step, individuals calculate a difference between the overall utilities associated with state + (adopting the innovation) and state - (non-adoption). For an agent $i$, the overall utilities of both states ($U_i^+$ and $U_i^-$) are defined by

$$U_i^\pm = \alpha_i \upsilon_i^\pm + (1-\alpha_i) u_i^\pm \ . \tag{5}$$

The expression for overall utility in Eq. (5) has two terms. The first term describes the contribution of social influence from decision-maker $i$'s social network, whereas the second term involves the contribution of agent $i$'s perceived utilities or preferences for states + or - , irrespective of other agents [42]. The social influence in Eq. (5) is quantified by $\upsilon_i^+$ or $\upsilon_i^-$, where $\upsilon_i^+$ is the proportion of social contacts of agent $i$ (those individuals to whom $i$ is connected by a first-degree social link, and assumed to have a social influence on $i$'s decisions) who already have adopted the innovation. Similarly, $\upsilon_i^-$ represents the proportion of social contacts who have not adopted yet. Clearly, both quantities must satisfy:

$$\upsilon_i^+ + \upsilon_i^- = 1 \ . \tag{6}$$

Agent $i$'s idiosyncratic preferences for an innovation are quantified by $u_i^+$ or $u_i^-$, where $u_i^+$ is the utility perceived by $i$ if he adopts the new product or technology. Conversely, $u_i^-$ represents the perceived utility of *not* adopting the innovation.

Parameter $\alpha_i \in 0,1$ is a measure of the relative weight of social influence and personal perceptions for agent $i$. The value of $\alpha_i$ may depend on the individual, but differences are also tied to the type of innovation considered [43]. For example, in fashionable markets (clothes, electronic gadgets or luxury items) social influence often has a large relative weight, whereas in other types of products (e.g., household appliances) social influence is smaller. Other factors that may affect the relative importance of social influence are the cost and complexity of the innovation, and the observability of results [44]. For simplicity, all our calculations assume that social influence and personal perception are weighed equally, that is, $\alpha_i = 0.5$.

Following [28], the probability of agent $i$ being in state + is 1, 0.5 or 0 depending on the sign of $\Delta U_i = U_i^+ - U_i^-$. If $\Delta U_i > 0$, agent $i$ will adopt the innovation (i.e., will be in state +) with probability 1; if $\Delta U_i = 0$, the probability of being in either state (+ or -) is 0.5; finally, if $\Delta U_i < 0$, agent $i$ will not adopt the innovation (i.e., probability of adoption is 0).

In order to compare the micro-model with the Bass model, a modification to our original ABM specification is necessary: here, we do not allow transitions from state + (adoption) to state - (non-adoption). This is because Bass' research was performed in the context of the consumer durables market, where disadoption is not observed. Therefore, the modified decision algorithm is as follows: if $\Delta U_i > 0$, agent *i* will irreversibly adopt the innovation; otherwise the agent remains in his current state.

## 3.2 Agent-based model implementation

In the Bass model there are two types of decision-makers: innovators and imitators. The ABM also involves two kinds of agents that are closely related (but not identical) to the Bass types. As in the Bass model, innovators in our ABM spontaneously adopt an innovation during the first few steps of the simulation and they are not influenced by social contagion. Unlike the Bass model, the adoption behavior of the rest of agents in our ABM does not only reflect social influence, but also individual preferences for an innovation.

Multiple software frameworks exist that reduce significantly the programming effort and time required to develop ABMs and the chances of making errors [45, 46]. We used REPAST Simphony, an open source framework maintained by Argonne National Laboratory [47].

The ABM environment is a 2-D lattice involving 40,000 nodes (a 200 x 200 grid). Each node represents an agent that makes decisions about the adoption of an innovation. Ties between nodes represent opportunities for contact between agents from the perspective of influencing adoption [27]. In the base model, agents are linked only to their nearest (first-order) neighbors in the lattice. All agents have the same number of social contacts (either 4 or 8, depending on the use of von Neumann or Moore neighborhoods) except for those located along the edges of the lattice, who have fewer contacts because periodic boundaries (i.e., a torus) are not used. In subsequent simulations, the regular network (the 2-D lattice) is replaced by alternative topologies built using the random rewiring procedure of [11] to generate small-world networks. This procedure is parameterized by a rewiring probability ($P_r$) that represents the probability of a first-order link being replaced by a "weak link" to a more distant node.

At each time step of the simulation, all agents who still have not adopted an innovation decide if they will adopt; otherwise agents remain in their previous state (remember, adoption is irreversible). As described above, the decision to adopt is based on the overall relative utility of the innovation. In turn, adoption is partly influenced by an agent's perception of the difference in utility between the new and old alternatives ( $\Delta u_i = u_i^+ - u_i^-$ ). The perceived value may be different for each individual, depending on his personal characteristics (e.g., his aversion to risk [48]) or his exposure to functional signals. For simplicity we assume here that the variability of personal perceptions among agents is small, thus justifying use of the same $\Delta u_i$ for all decision makers. Moreover, $\Delta u_i$ is assumed to be constant in time. In a more realistic scenario, an inhomogeneous distribution of $\Delta u_i$ values should be considered. A simulation continues until only small fluctuations are observed in the adoption pattern, or a given option prevails completely.

## 3.3 Introduction of innovators

In the Bass model, the number of innovators grows asymptotically with time. To reflect this characteristic, innovators are exogenously introduced into the ABM at the beginning of the simulation according to a heuristic procedure based on Eq. (1). Eq. (1) is rewritten in integral form, replacing $n^+$ with $N^+/N$, where $N^+$ is the number of adopters and $N$ is the total number of agents in the population:

$$N^+(t) = p \int_0^t \left[N - N^+(t')\right] dt' + q \int_0^t \frac{N^+(t')}{N} \left[N - N^+(t')\right] dt' + N_0^+ , \qquad (7)$$

where $N_0^+ \equiv N^+(0)$. In the following, to be consistent with Eq. (2), at the start of a simulation all individuals are assumed to be in state -; that is, $N_0^+ = 0$.

Eq. (7) involves two terms, respectively multiplied by parameters $p$ and $q$ of the Bass model. The first term represents the contribution of innovators to the total number of adopters, whereas the second term denotes the contribution of imitators. Then,

$$N_I(t) \equiv p \int_0^t \left[N - N^+(t')\right] dt' . \qquad (8)$$

In order to define a first-order approximation to the contribution of innovators, the time derivative of $N_I(t)$ at $t = 0$ is calculated; this value is called $\gamma$ and describes the rate of introduction of innovators into the ABM. Then, from the derivation of Eq. (8) we get

$$\gamma = pN . \qquad (9)$$

The introduction of innovators does not continue throughout the simulation, but only until a final proportion of 2.5% of innovators is reached, following Rogers' definition for this type of adopters [49]. As the population size in our ABM is 40,000, 1000 innovators need to be introduced, but they can be added at different rates. From Eq. (9), and considering a plausible range of values for $p$ = [0.003, 0.025], a set of $\gamma$ values can be calculated for use in different simulations. In subsequent experiments, we use values of $\gamma$ = [125, 200, 250, 500, 1000] innovators per simulation step.

# 4 Results

## 4.1 Overview

The central motivation behind the experiments presented below is to explore associations between microscopic parameters in an ABM of innovation diffusion and parameters of the macroscopic Bass model. First we simulate multiple cumulative adoption trajectories over time by varying the most important micro-parameters in the ABM (Section 4.2). Then, we estimate values of $p$ and $q$ Bass parameters and calculate takeoff time for each simulated adoption trajectory (Section 4.3). Next, we explore how $p$ and $q$ values respond to changes in ABM micro-variables (Section 4.4). The Bass parameters derived from micro-level simulations are compared to values reported in the literature for actual innovations (Section 4.5). We investigate associations between estimated takeoff times and ABM micro-variables (Section 4.6). Finally, we present a simple example (Section 4.7) to illustrate possible use of results.

## 4.2 Generation of synthetic adoption trajectories using a microscopic model

In this section, we generate multiple simulated adoption trajectories by varying the values of important parameters in our micro-level ABM. The microscopic parameters considered can be grouped into two major categories that respectively represent: (a) external influences linked to innovation processes ($\Delta u$, $\gamma$, and $\sigma$, described below) and (b) internal influences related to imitation processes ($P_r$ and $k$).

Parameter $\Delta u$ describes the difference in utilities between adoption and non-adoption (or alternatively, new and existing products or technologies). As described above, this variable reflects an individual's perception that can be influenced by external factors such as improvements in the performance or decreases in the price of an innovation. Two values will be explored for this variable, 0.6 and 0.8 that respectively indicate positive and highly positive utilities for the simulated innovation. These values ensure that market saturation (100% adoption) is reached.

The second parameter considered is the rate of introduction of innovators $\gamma$. This parameter can plausibly be tied to advertising – among other external influences – but not to word of mouth. As described above, innovators (here, 1000 or 2.5% of the population) can be introduced into the simulations at different rates (see Section 4.1). As all simulations take from 15 to 45 steps to reach full adoption, we chose 8 steps as the maximum number of steps during which innovators are introduced; this is because in the Bass model innovation occurs in the earlier stages of the adoption process. We specify, then, that adoption by innovators can take place in 8, 5, 4, 2 or 1 simulation steps or "ticks," corresponding to rates of 125, 200, 250, 500 and 1,000 adoptions per tick, respectively.

Because we are simulating adoption on a lattice that reflects social distance among agents, the distribution of innovators can have an influence on subsequent diffusion [50]. We explore three different patterns of spatial dispersion ($\sigma$), ranging from a compact group of innovators in the center of the lattice to a uniform distribution of innovators. An intermediate pattern involves a few clusters of innovators throughout the lattice. The distribution of innovators loses importance as the network in which diffusion takes place changes from regular to fully random topology.

The second group of micro-model parameters is related to the topology of the market's social network. We explore two of these parameters. The first one is the probability of rewiring between adjacent nodes, $P_r$. The lowest value considered is $P_r$=0, that corresponds to a completely regular network; other $P_r$ values considered are 0.0025, 0.005, 0.01, 0.02, and 0.04. These values are intended to span the full range of values corresponding to small-world networks [51].

The final parameter to be explored is $k$ – the degree of the social network, i.e., the average number of social contacts for each agent. Individuals who are related to many others who already have adopted an innovation may have greater probability of adoption because their peers can provide information about the advantages or risks of the innovation and ultimately exercise greater joint influential power. This has been referred to as the "degree effect," after the number of connections (degree) an individual has. We simulate two network degree values: 4 and 8. These values correspond to 4 and 8 first-order neighbors in a regular network (i.e., von Neumann and Moore neighborhoods in a spatial lattice). For $P_r > 0$, although the regular lattice is rewired, the number of contacts for each agent is modified but the average degree of the network remains unchanged.

The different values considered for the five parameters in our micro-model yield 360 unique combinations: we simulated adoption for each of these combinations. These simulations resulted in 360 synthetic adoption trajectories (i.e., the cumulative proportion of adopters as a function of time).

## 4.3   Fitting parameters of the Bass model

Parameters of the Bass model can be estimated from empirical data using a variety of methods such as ordinary least squares [1], maximum likelihood [52], nonlinear least squares [53] and genetic algorithms [54]. We estimated values of $p$ and $q$ for each of the 360 simulated adoption trajectories described in the previous section using function NonLinearModelFit in Mathematica 8 [55] with default parameters. A few illustrative fits are shown in Fig. 1. All four panels correspond to simulations using $\Delta u$ = 0.6, $k$ = 8 and a "uniform" distribution of innovators ($\sigma$). The top two panels (Figs. 1a and 1b) correspond to a fully regular network ($P_r$ = 0) and have $\gamma$ values of 200 and 500 innovators per step (i.e., innovators included in 5 and 2 steps respectively). The bottom panels (Figs. 1c and 1d) correspond to the upper end of small-world networks ($P_r$ = 0.04, where characteristic network path lengths are low and clustering coefficients are high) and the same values of 200 and 500 as the top panels. Notice that the number of simulation steps required to reach market saturation (the x-axis) is larger in the top panels, reflecting faster adoption in small-world networks. The fitted Bass models closely match the results of micro-model realizations under all combinations of micro-parameters considered. For all simulated trajectories (including those not shown on Fig. 1), the parameter fits for the Bass model are very good: in all cases $r^2$ values are very high (> 0.98); results from all fits are show on Table 1 in the supplementary information accompanying the electronic version of this paper.

[ Insert Fig. 1 here ]

*Fig. 1: Simulated adoption trajectories and induced values of Bass model parameters. The adoption trajectories shown were simulated with common micro-parameters $\Delta u$ = 0.6, $k$ = 8 and a uniform distribution of innovators ($\sigma$). Panels (a) and (b) correspond to values of $\gamma$ = 200 and 500 respectively for a regular network. Panels (c) and (d) correspond to the same values used in (a) and (b) respectively, but for a rewiring probability of 0.04.*

## 4.4   Linking p and q Bass parameters to changes in microscopic parameters

Estimated $p$ and $q$ values can be associated with the parameters of the microscopic model used to simulate adoption patterns. Fig. 2 shows the boundaries of a region that encompasses all values of $p$ and $q$ estimated for adoption trajectories simulated with a subset of combinations of micro-parameters: $\Delta u$ = 0.6, $k$ = 8, and a "compact" spatial distribution of innovators. This region spans all $p$ and $q$ values induced by the simulated adoptions. An important implication is that any $p-q$ combination within this region can be mapped to a particular combination of micro-model parameters. Thus, we can assess aggregate adoption patterns resulting from plausible interventions via changes in micro-parameters.

[ Insert Fig. 2 here ]

*Fig. 2: Variation in Bass parameters $p$ and $q$ in response to changes in parameters of the microscopic model. The region displayed is the outer envelope of $p$ and $q$ values for adoption trajectories simulated with a subset of combinations of micro-parameters: $\Delta u$ = 0.6, $k$ = 8, and a "compact" spatial distribution of innovators. The grey arrows along the region boundaries indicate the direction of variation in micro-parameters $\gamma$ and $P_r$.*

The boundaries of the region shown in Fig. 2 are defined by combinations of two micro-model parameters: five $\gamma$ values (from 125 to 1000 innovators per tick) and six $P_r$ values (from 0.00 to 0.04). The grey arrows indicate how parameters vary along the boundaries. Although we simulated three different spatial distributions of innovators, this parameter did not have a substantial effect on $p$ and $q$, so for clarity of presentation we only show results for one $\sigma$ ("compact"). The upper and lower boundaries of the region in Fig. 2 are associated with variation of the rate of introduction of innovators into a simulation. From left to right, $\gamma$ increases from 125 to 1000 innovators per tick. That is, as innovators are introduced earlier, the induced values of $p$ increase. This behavior seems plausible because $p$ is tied to innovation processes and has greater relevance during the early stages of an adoption. At the same time, variations in $\gamma$ do not induce substantial changes in the $q$ parameter (the y-axis in Fig. 2) and therefore the upper and lower boundaries are relatively horizontal. It seems reasonable to expect that changes in numbers of early adopters will not influence the probability of imitation if the linkage between micro- and macroscopic models is not only mathematical but also conceptual.

The two sides of the region in Fig. 2 are defined by $p$ and $q$ values induced as the topology of the social network in the micro-model changes from a regular network ($P_r$ = 0.00) at the bottom of the region to a SWN with $P_r$ = 0.04 at the top. That is, the induced changes in $q$ result from the full range of rewiring probabilities associated with small-world networks. As we move up along the side boundaries, the values of $q$ more than double from 0.35 to about 0.70. These changes in $q$ are reasonable, because an increase in the probability of rewiring decreases the average path length of the network, which in turn enhances linkages among nodes that facilitate imitation processes. Differences in network topology also seem to induce changes (a slight decrease) in $p$; that is, the side boundaries of the region are oblique, not vertical, because $p$ also changes in response to variations in $P_r$. The reasons for the induced changes in $p$ are unclear, as one would not expect the network topology to influence innovation, a process assumed to respond solely to external influences. Following the probabilistic line of thought that inspired the Bass model, perhaps parameter $p$ should not be conceptualized as a spontaneous generation of adopters totally isolated from social context, as this seems unlikely. On the contrary, innovating behavior can be influenced by an agent's social identity, in turn somehow tied to his social network. For example, if a farmer perceives that "being a good farmer" involves adopting new technologies earlier than peers, then innovating characteristics may be tied to an agent's self-perceived social role within his network. Such a linkage might explain the dependence of $p$ on $P_r$.

A region similar to that in Fig. 2 can be defined for a network of degree $k$ = 4 (all other micro-parameters remained the same). For the sake of space, this region is shown in Fig. 3 (in dashed lines). The shape of the region is similar to that for $k$ = 8, and the same considerations apply to the ties between micro-model parameters ($\gamma$ and $P_r$) and induced $p$ and $q$ values. The lower average network degree, however, displaces the region towards higher $p$ values (and, to a lesser extent, towards higher $q$ values). The reason for this displacement is not immediately intuitive, but is tied to changes in the social influence necessary (represented by the number of adopter neighbors) to cause adoption for a given relative utility. In our adoption model, a reduction in the degree of the network is equivalent to a reduction in the number of adopter neighbors required for an agent to adopt. For instance, for $k$ = 8 and $\Delta u$ = 0.6, two adopter neighbors are necessary to trigger adoption, whereas for $\Delta u$ = 0.8 (i.e., higher perceived benefits) only one adopter neighbor is needed. In contrast, for $k$ = 4 only one adopter neighbor is needed for $\Delta u$ values of both 0.6 and 0.8 (for this reason, results for $\Delta u$ = 0.8 were not plotted).

## 4.5 Comparison of simulated and real-world *p* and *q* values

Sixty-eight values of $p$ and $q$ from actual adoption patterns reported in five published studies [1, 29-31, 33] are shown in Fig. 3. We plot here the region shown in Fig. 2 (in a solid line) and display a second region (in dashed lines) corresponding to the envelope for a second subset of simulations with a network of degree $k$ = 4 (see discussion in previous section).

[ Insert Fig. 3 here ]

*Fig. 3: Comparison of p and q values estimated from actual adoption processes and those induced by simulated adoption trajectories. The region bound by the solid lines corresponds to that shown in Fig. 2 (k = 8). The region indicated with dashed lines corresponds to $k = 4$; all other combinations of micro-parameters remain as for the solid-line region.*

Many of the $p$ and $q$ values reported in the literature are within or near the boundaries of the two regions that encompass our simulated values (Fig. 3). The observed points correspond mostly to consumer goods in the U.S. and Spain [1, 29, 30], and to Internet use, broadband adoption and cell phone use in the U.S. and Argentina [33]. In contrast, other $p$ and $q$ values are not spanned by the two regions. The points outside the simulated regions correspond mostly to consumer goods introduced in Europe [31]. The location of these points in $p-q$ space suggests a greater importance of innovation processes. This might be related to the fact that these products had been released in the U.S. prior to their introduction in Europe. European users, therefore, may have been exposed to prior information about these products. On the other hand, there may be cultural differences in adoption patterns [9, 40, 56].

The $q/p$ ratio summarizes the shape of a cumulative adoption curve and can be interpreted as a shape parameter [24]. The points not contained within the two regions in Fig. 3 share a common characteristic: their $q/p$ ratio is lower than for points that are within or near the two regions. Reference [9] describes the effect of many individual and social behaviors that may influence the $q/p$ value. For example, individualism and heterogeneity are factors that tend to lower this value. Individualism can be modeled by means of a weighted balance between the individual preference and the social pressure. In our ABM, this is mathematically equivalent to increasing the value of $\Delta u$, a change that has already been analyzed. Heterogeneity, in turn, requires use of distributions of individual characteristics for decision-makers. This approach has not been considered here, but should be addressed in future work.

### 4.6 Simulated takeoff times as a function of micro-parameters

In this section we explore changes in $t_{TO}$ in response to $\gamma$ and $P_r$, the two micro-level parameters that appeared to influence induced values of Bass parameters. Figs. 4a and 4c display changes in $t_{TO}$ estimated from the ABM simulations as a function of the rate of introduction of innovators. These panels correspond to networks of degree 8 and 4, respectively. The various lines represent different combinations of micro-parameters, but we only highlight those quantities that introduce noticeable differences in the results. Fig. 4a shows that $t_{TO}$ decreases as $\gamma$ increases. Faster introduction of innovators (i.e., higher $\gamma$) accelerates clustering of adopters and adoption thresholds are reached earlier, thus reducing time to takeoff.

The other important variable in Fig. 4 is $\Delta u$. Two clusters of lines are apparent in Fig.4a that correspond to $\Delta u$ values of 0.6 (top) and 0.8 (bottom) respectively. When $\Delta u$ is increased from 0.6 to 0.8 (i.e., the relative advantage of the innovation is greater), $t_{TO}$ decreases noticeably. Changes in network topology and dispersion of innovators (that give rise to the various lines in Fig. 4) seem to have second-order effects, as they do not modify noticeably the values of $t_{TO}$. Fig. 4c shows the same as Fig. 4a, but for a network with fewer neighbors ($k$ = 4). As discussed in Section 4.4, with smaller number of neighbors both values of $\Delta u$ have the same threshold of social influence (one adopter neighbor) required to trigger adoption. Consequently, the effect of $\Delta u$ disappears and the previous two clusters of lines in Fig. 4a now are merged into a single cluster (most lines overlap and cannot be discerned in the figure).

Figs. 4b and 4d illustrate changes in $t_{TO}$ as a function of the probability of network rewiring; the panels correspond to networks degrees of 8 and 4, respectively. Fig. 4b confirms that $P_r$ does not seem to have much influence on $t_{TO}$, as all lines are relatively horizontal. Instead, changes in $t_{TO}$ continue to be influenced by $\Delta u$ and $\gamma$. For $\Delta u$ = 0.6, $t_{TO}$ decreases when $\gamma$ is increased from 125 to 1000. The curves shift to lower $t_{TO}$ values when $\Delta u$ increases to 0.8. Nevertheless, the effect of $\gamma$ persists, as the curve for $\gamma$ = 1000 is below the one for $\gamma$ = 125. Fig. 4d corresponds to the network with 4 neighbors. As in Fig. 4c, now the two $\Delta u$ values have the same threshold of adoption (one adopting neighbor). Consequently, the effect of $\Delta u$ disappears and the various curves separate only by their $\gamma$ value.

[ Insert Fig. 4 here ]

*Fig. 4: Takeoff time as a function of variations in parameters $\gamma$ (panels a and c) and Pr (panels b and d). Top panels correspond to simulations with a network of degree k =8. Bottom panels are for k=4.*

## 4.7  An example application

To illustrate the possible application of our approach, we present a hypothetical but plausible example. We focus on a multinational corporation that is introducing a new product in different markets (e.g., different countries). We assume the innovation has already been introduced in a few markets, one of which ($M_0$) has similar socio-economic characteristics to the new market (M) where the product will be introduced. Moreover, we assume the previous and new markets are isolated from one another (i.e., there is no interaction among decision-makers in both markets). The company wishes to enhance the cumulative profits from the innovation (between introduction and a specific time) via investment in advertising. To achieve this goal, the company may follow the steps described below.

1. We assume that the product's diffusion in market M will be relatively similar to that observed in $M_0$ [57]. This implies that, under the same circumstances (i.e., initial conditions and parameters), the values of Bass parameters for market M $(p,q)$ are equal to those that characterize the previous diffusion pattern in $M_0$ $(p_0,q_0)$.

2. We assume that the diffusion pattern defined by $(p_0,q_0)$ falls within one of the two regions of coincidence between micro and macro models (Fig. 3). Then, we can find the combination of micro-parameters that induced the diffusion described by $(p_0,q_0)$.

3. In order to assess the outcome of manipulating control variables to enhance diffusion and profits, an association should be established between a plausible control variable (e.g., an additional investment $I$) and relevant micro-parameters. In this example we focus on micro-parameter $\gamma$ – the rate of introduction of innovators. Defining the shape of the association is beyond the scope of this paper and should be the object of marketing research. Nevertheless, without loss of generality, we can assume that micro-parameter $\gamma$ is an increasing function of investment, i.e., $\gamma = \gamma(I)$.

4. Another variable that needs to be defined is the time $t^*$ at which the cumulative profits from the innovation will be assessed. We assume that $t^* > t_{TO}$ for any adoption pattern, as it would be unreasonable to expect to recover investments before an innovation has taken off (at time $t_{TO}$).

5. As shown in Fig. 4, $\gamma_0 < \gamma_I \Rightarrow t_{TO}(\gamma_0) > t_{TO}(\gamma_I)$; the subindices $I$ and $0$ indicate the adoption processes with and without additional investment respectively. It can be shown through analysis of the cumulative adoption curves that $n^+[\gamma_1, t^*] > n^+[\gamma_0, t^*]$ for any time $t^* > t_{TO}$. However, we must determine the impact on cumulative net profits from increased sales.

6. We define G as the cumulative net profit after subtracting the additional investment in advertising. Then, it is reasonable to assume that G is a known function of $n^+[\gamma, t^*]$, which in turn is parameterized by $\gamma$ and $t^*$. The described procedure allows the company to assess whether the expected profits G are sufficient to recover the investment in advertising and derive a desired return on this investment. In other words, the company may assess whether

$$G[\gamma_1, t^*] - G[\gamma_0, t^*] > ROI_{min} , \qquad (10)$$

where $ROI_{min}$ is the minimum desired return on investment at time $t^*$. Eq. 10 defines the criterion to evaluate whether the additional inversion in advertising is profitable or not.

## 5 Conclusions

We have successfully aligned [58] micro- and macro-level approaches to modeling diffusion of innovations. Several micro-level simulations performed with different parameter combinations match very closely the aggregate adoption patterns predicted by the widely-used Bass model. Moreover we showed that, at least for a portion of the domain, results from micro-simulations are consistent with observed aggregate-level adoption patterns. Specifically, combinations of $p$ and $q$ Bass parameters estimated from actual adoption trajectories can be induced by the processes specified in our micro-level model of innovation adoption. The main implication is that any $p-q$ combination within the domain of coincidence (the two regions shown in Fig. 3) can be tied to a particular combination of micro-model parameters. In turn, this allows us to explore changes in macro adoption patterns resulting from plausible variation in micro-parameters.

We explored how $p$ and $q$ respond to changes in a few micro-parameters. Major results are as follows: (1) $p$ increased not only with the number of innovators, but also with the rate at which innovators were introduced; (2) $q$ increased with the probability of rewiring in small-world networks, as the characteristic path length decreased; and (3) an increase in the overall perceived utility associated with an innovation caused a corresponding increase in induced $p$ and $q$ values.

The takeoff time $t_{TO}$ is another useful macro-level indicator of the penetration of an innovation. Because takeoff time is a function of $p$ and $q$, we also established an association between $t_{TO}$ and micro-level parameters. We found that the takeoff time was strongly affected by three variables: the difference in utilities between adoption and non-adoption ($\Delta u$), the rate of introduction of innovators ($\gamma$) and the average degree ($k$) of the social network underlying diffusion. Understanding which factors accelerate the time to take-off has important practical implications, as fast adoption requires appropriate production, distribution and marketing strategies that often require a considerable amount of investment and lead time [5].

Macroscopic models provide parsimonious and analytically tractable descriptions of innovation diffusion. Nevertheless, this type of models does not provide insight about the individual processes and social interactions that may drive adoption [5]. In contrast, microscopic models easily capture the networks of relationships among individuals and heterogeneity in their attributes, but they increase computational requirements, constrain sensitivity analysis, and demand greater decision-maker time and attention. By contrasting micro and macro models of innovation diffusion, we can assess the importance of relaxing some of the assumptions (e.g., the perfect mixing and homogeneity assumptions) in the simpler macro models [26]. The alignment of micro- and macro-level modeling approaches does not seek to invalidate the widely-used Bass model, but to gain insight – within the domain of coincidence – of plausible processes underlying macroscopic patterns.  An analogy is how Statistical Mechanics does not invalidate Thermodynamics (that allows useful observation of relevant quantities), but attempts to explain its microscopic underpinnings.

From an applied point of view, understanding micro to macro linkages might allow marketing experts to design interventions on micro-variables – or processes related to them – that might enhance adoption of future products or technologies. An example of a possible application of our approach was presented in Section 4.7. Interventions such as lowering the introductory price of an innovation can help to jump-start spontaneous adoption by innovators. Another plausible strategy to enhance adoption by innovators may involve heavy initial investment in advertising to "prime the word of mouth pump," subsequently reducing the level of spending as the product diffuses (but see [59]). Alternatively, the growing number of low-cost marketing technologies and media available for triggering social influence makes low advertising budgets at launch increasingly feasible. Micro-model parameters may be set to mimic individual effects of alternative marketing activities, and plausible individual-level responses can be included in the decision rules of agents [5]. Simulated aggregate results can then be used to assess the performance of different marketing strategies over the aggregate population.

# 6   Acknowledgements

This research was supported by cooperative agreement SES-0951516 from the U.S. National Science Foundation's Decision Making Under Uncertainty (DMUU) program to the Center for Research on Environmental Decisions (CRED). Additional support was provided by NSF program "Decadal and Regional Climate Prediction using Earth System Models (EaSM)", grant 104910, and by the University of Buenos Aires.

Journal of Marketing, 73 (2009) 1-13.
[23] J. Goldenberg, B. Libai, E. Muller, The chilling effects of network externalities: Perspectives and conclusions, International Journal of Research in Marketing, 27 (2010) 22-24.
[24] R. Chatterjee, J. Eliasberg, The Innovation Diffusion Process in a Heterogeneous Population: A Micromodeling approach, Management Science, 36 (1990) 1057-1079.
[25] J. Goldenberg, S. Efroni, Using cellular automata modeling of the emergence of innovations, Technological Forecasting and Social Change, 68 (2001) 293-308.
[26] H. Rahmandad, J. Sterman, Heterogeneity and Network Structure in the Dynamics of Diffusion: Comparing Agent-Based and Differential Equation Models, Management Science, 54 (2008) 998-1014.
[27] N.I. Shaikh, A. Rangaswamy, A. Balakrishnan, Modeling the diffusion of innovations through small-world networks, in, Electronic copy available at: http://ssrn.com/abstract=2032861, 2010.
[28] C.E. Laciana, S.L. Rovere, Ising-like agent-based technology diffusion model: Adoption patterns vs. seeding strategies, Physica A: Statistical Mechanics and its Applications, 390 (2011) 1139-1149.
[29] F.M. Bass, T.V. Krishnan, D.C. Jain, Why the Bass Model Fits Without Decision Variables, Marketing Science, 13 (1994) 203-223.
[30] I.R. Bellón, I.C. Roche, El Proceso de Difusión de Nuevos Productos: Aplicación al Microondas y Compact Disc en España, Estudios sobre consumo, (2000).
[31] H. Gatignon, J. Eliashberg, T.S. Robertson, Modeling Multinational Diffusion Patterns: An Efficient Methodology, Marketing Science, 8 (1989) 231-247.
[32] F. Sultan, J. Farley, D. Lehmann, A Meta-Analysis of Applications of Diffusion Models, Journal of Marketing Research, (1990) 70-77.
[33] V. Weissmann, Difusión de Nuevas Tecnologías y Estimación de la Demanda de Nuevos Productos: un Análisis Comparativo entre Argentina y EE.UU., Palermo Business Review, (2008) 5-17.
[34] R.B. Cialdini, N.J. Goldstein, Social Influence: Compliance and Conformity, in: Annual Review of Psychology, 2004, pp. 591-621.
[35] R. Guseo, M. Guidolin, Modelling a Dynamic Market Potential: A Class of Automata Networks for Diffusion of Innovations, in: Technological Forecasting and Social Change, 2009, pp. 806-820.
[36] P.N. Golder, G.J. Tellis, Will It Ever Fly? Modeling the Takeoff of Really New Consumer Durables, Marketing Science, 16 (1997) 256-270.
[37] B.L. Lim, M. Choi, M.C. Park, The Late Take-Off Phenomenon in the Diffusion of Telecommunication Services: Network Effect and the Critical Mass, Information Economics and Policy, 15 (2003) 537-557.
[38] P.N. Golder, G.J. Tellis, Growing, Growing, Gone: Cascades, Diffusion, and Turning Points in the Product Life Cycle, Marketing Science, 23 (2004) 207-218.
[39] V. Mahajan, E. Muller, Innovation Diffusion and New Product Growth Models in Marketing, Journal of Marketing, (1979) 55-68.
[40] G.J. Tellis, S. Stremersch, E. Yin, The International Takeoff of New Products: The Role of Economics, Culture, and Country Innovativeness, Marketing Science, 22 (2003) 188-208.
[41] E. Ising, Beitrag zur Theorie des Ferromagnetismus, Zeitschrift für Physik, 31 (1925) 253-258.
[42] G. Weisbuch, G. Boudjema, Dynamical aspects in the adoption of agri-environmental measures, Advances in Complex Systems, 2 (1999) 11-36.
[43] S.A. Delre, W. Jager, T.H.A. Bijmolt, M.A. Janssen, Targeting and timing promotional activities: An agent-based model for the takeoff of new products, Journal of Business Research, 60 (2007)

Figure 1

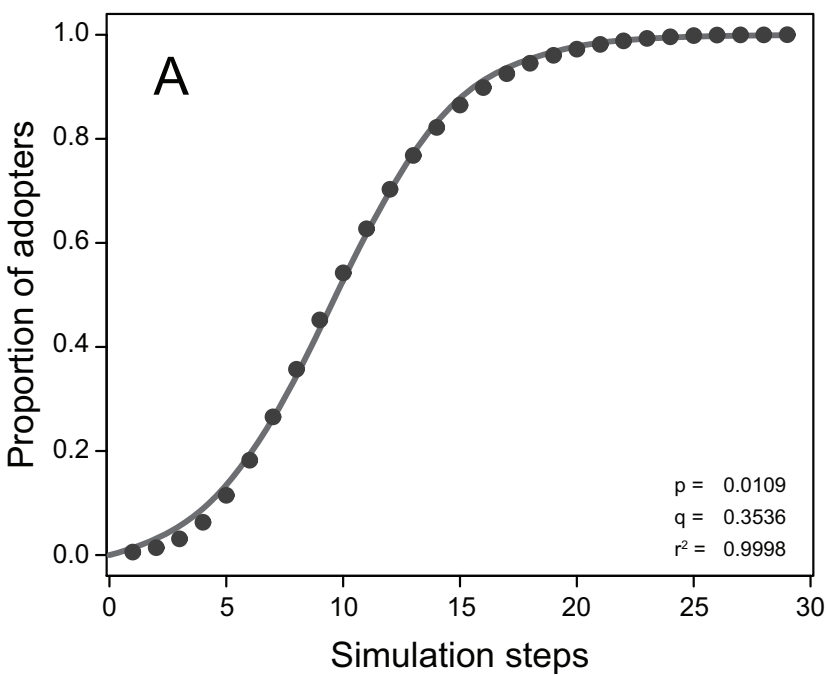
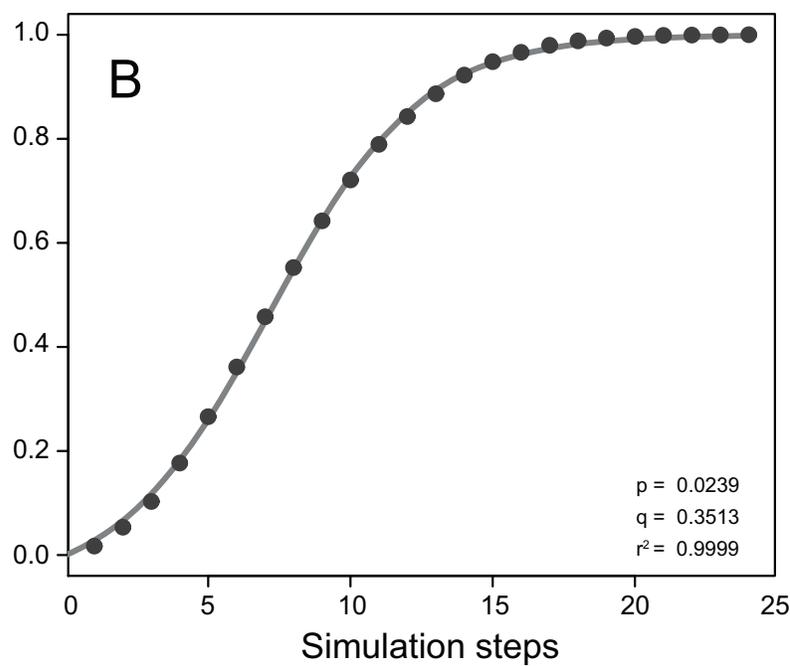
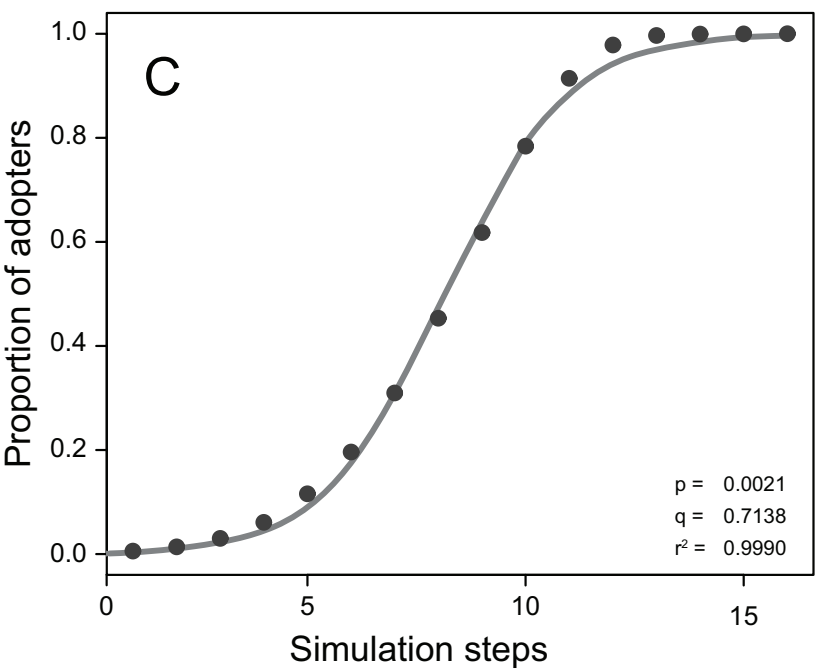
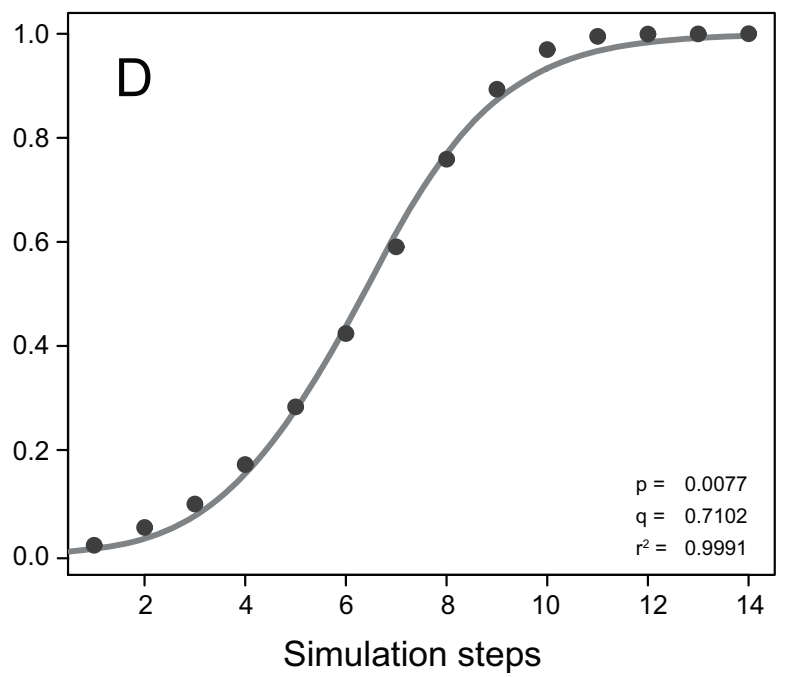

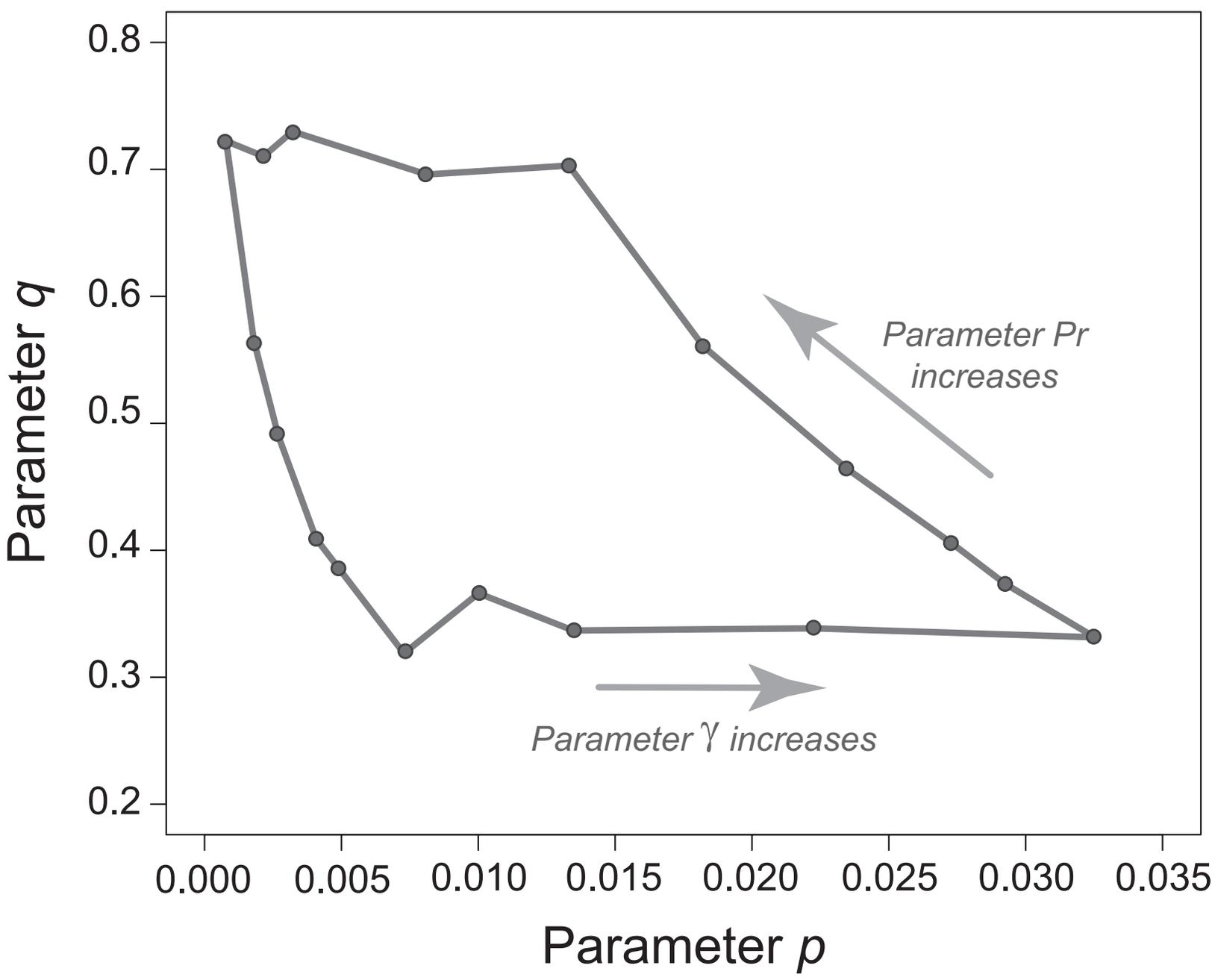

Figure 3

Figure 4

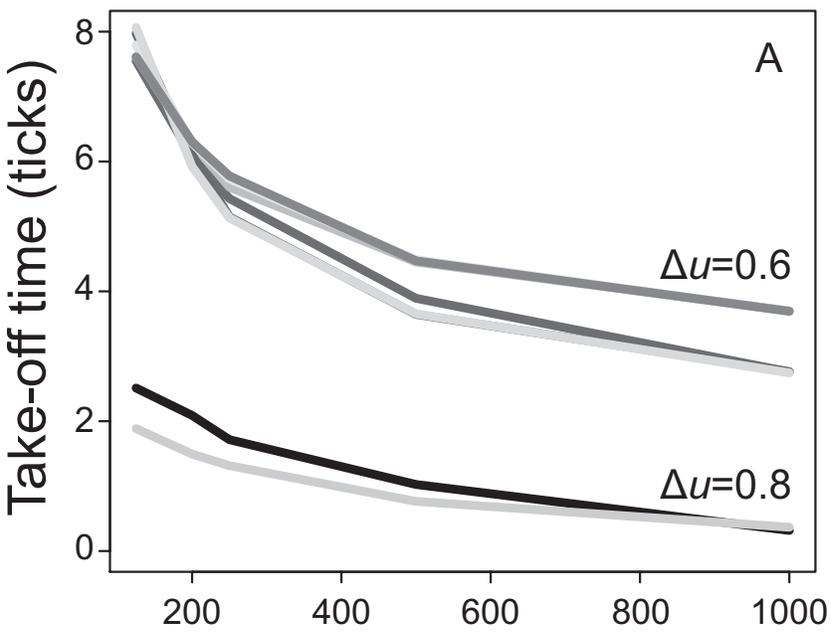
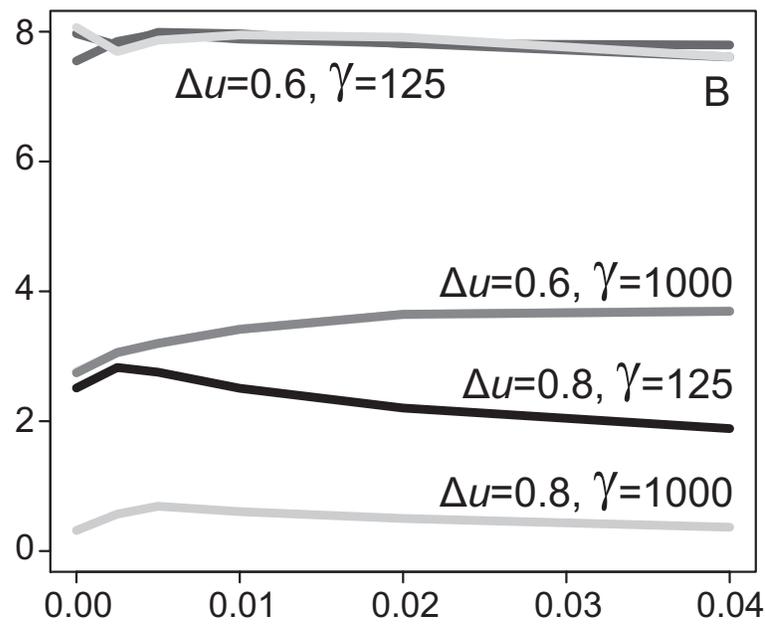
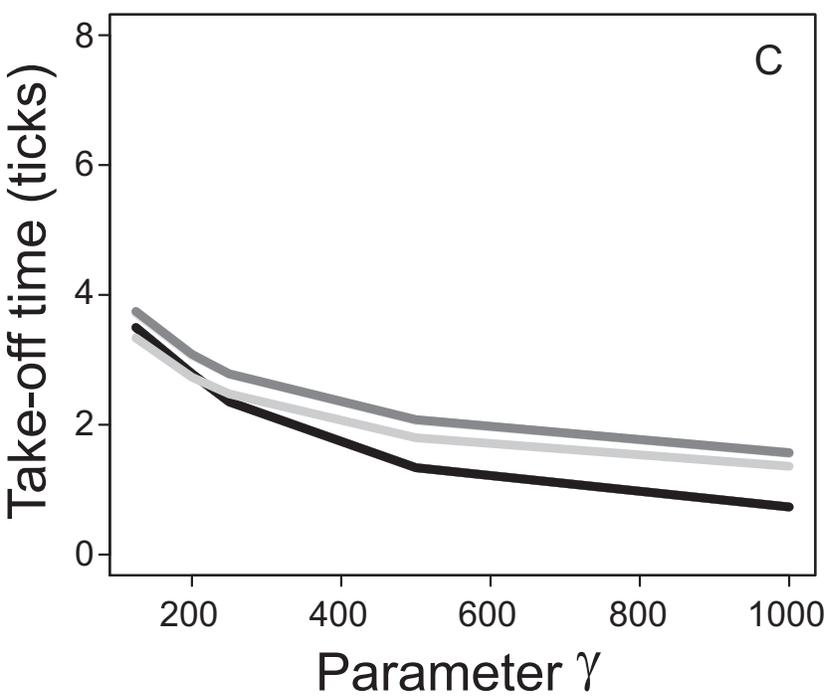
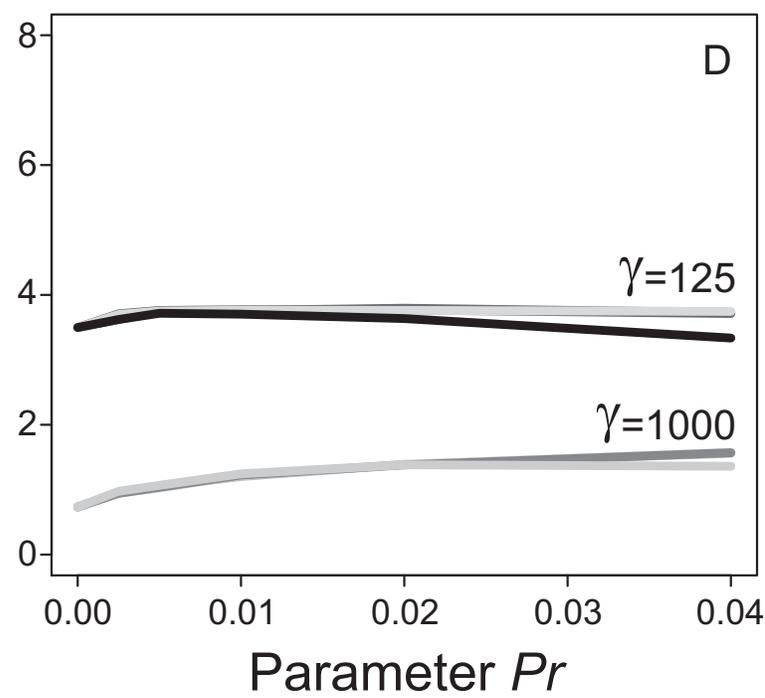

Exploring associations between micro-level models of innovation diffusion
and emerging macro-level adoption patterns (Supplementary Data)

| k | Δμ | σ | Pr | γ | p | q | R² | Take-off time |
|---|---|---|---|---|---|---|---|---|
| 8 | 0,6 | COMPACT | 0 | 125 | 0,0072863 | 0,3187899 | 0,9998206 | 7,549109451 |
| 8 | 0,6 | COMPACT | 0 | 200 | 0,010011 | 0,3660977 | 0,9998548 | 6,068078523 |
| 8 | 0,6 | COMPACT | 0 | 250 | 0,0134703 | 0,3368193 | 0,9998985 | 5,430077886 |
| 8 | 0,6 | COMPACT | 0 | 500 | 0,0222681 | 0,3385364 | 0,9999436 | 3,892737311 |
| 8 | 0,6 | COMPACT | 0 | 1000 | 0,0325243 | 0,3314418 | 0,9999571 | 2,759889255 |
| 8 | 0,6 | COMPACT | 0,0025 | 125 | 0,0047668 | 0,3878899 | 0,9999136 | 7,84932127 |
| 8 | 0,6 | COMPACT | 0,0025 | 200 | 0,0079722 | 0,4102945 | 0,9998418 | 6,273391582 |
| 8 | 0,6 | COMPACT | 0,0025 | 250 | 0,0109441 | 0,3926135 | 0,999943 | 5,607803453 |
| 8 | 0,6 | COMPACT | 0,0025 | 500 | 0,0224628 | 0,3529985 | 0,9999647 | 3,829011972 |
| 8 | 0,6 | COMPACT | 0,0025 | 1000 | 0,0292511 | 0,3735879 | 0,9998947 | 3,054017495 |
| 8 | 0,6 | COMPACT | 0,005 | 125 | 0,0040758 | 0,4090616 | 0,9999415 | 7,96789295 |
| 8 | 0,6 | COMPACT | 0,005 | 200 | 0,0072751 | 0,4433309 | 0,9999229 | 6,19811199 |
| 8 | 0,6 | COMPACT | 0,005 | 250 | 0,0091892 | 0,4511654 | 0,9998761 | 5,597524233 |
| 8 | 0,6 | COMPACT | 0,005 | 500 | 0,0205429 | 0,3888078 | 0,9999701 | 3,966309529 |
| 8 | 0,6 | COMPACT | 0,005 | 1000 | 0,0272767 | 0,405882 | 0,9999359 | 3,192987123 |
| 8 | 0,6 | COMPACT | 0,01 | 125 | 0,0026925 | 0,4903028 | 0,9998841 | 7,885680144 |
| 8 | 0,6 | COMPACT | 0,01 | 200 | 0,0053541 | 0,4936355 | 0,9998824 | 6,426948778 |
| 8 | 0,6 | COMPACT | 0,01 | 250 | 0,0065528 | 0,5136461 | 0,9998822 | 5,85293821 |
| 8 | 0,6 | COMPACT | 0,01 | 500 | 0,0163694 | 0,4768216 | 0,9998753 | 4,166282945 |
| 8 | 0,6 | COMPACT | 0,01 | 1000 | 0,02348 | 0,4640829 | 0,9998124 | 3,418958649 |
| 8 | 0,6 | COMPACT | 0,02 | 125 | 0,0018077 | 0,5640512 | 0,9995902 | 7,821938222 |
| 8 | 0,6 | COMPACT | 0,02 | 200 | 0,0036931 | 0,5945183 | 0,9996329 | 6,29264336 |
| 8 | 0,6 | COMPACT | 0,02 | 250 | 0,0052598 | 0,5864115 | 0,9997348 | 5,741312288 |
| 8 | 0,6 | COMPACT | 0,02 | 500 | 0,0106112 | 0,6025919 | 0,9997842 | 4,439595198 |
| 8 | 0,6 | COMPACT | 0,02 | 1000 | 0,0182224 | 0,5606277 | 0,9996221 | 3,644202156 |
| 8 | 0,6 | COMPACT | 0,04 | 125 | 0,0007887 | 0,7223214 | 0,9986266 | 7,610104093 |
| 8 | 0,6 | COMPACT | 0,04 | 200 | 0,0021542 | 0,7102671 | 0,9990325 | 6,290173749 |
| 8 | 0,6 | COMPACT | 0,04 | 250 | 0,0032137 | 0,7298703 | 0,99915 | 5,604387399 |
| 8 | 0,6 | COMPACT | 0,04 | 500 | 0,0081198 | 0,6959686 | 0,9991647 | 4,451211041 |
| 8 | 0,6 | COMPACT | 0,04 | 1000 | 0,0133122 | 0,7029824 | 0,9986852 | 3,699169256 |
| 8 | 0,6 | INTERMEDIATE | 0 | 125 | 0,004761 | 0,3792734 | 0,9999155 | 7,970210528 |
| 8 | 0,6 | INTERMEDIATE | 0 | 200 | 0,0105125 | 0,3464453 | 0,999833 | 6,102122243 |
| 8 | 0,6 | INTERMEDIATE | 0 | 250 | 0,014986 | 0,3297194 | 0,9997331 | 5,146919968 |
| 8 | 0,6 | INTERMEDIATE | 0 | 500 | 0,023966 | 0,351264 | 0,9999177 | 3,645614138 |
| 8 | 0,6 | INTERMEDIATE | 0 | 1000 | 0,0324967 | 0,331369 | 0,9999582 | 2,762385728 |
| 8 | 0,6 | INTERMEDIATE | 0,0025 | 125 | 0,0046062 | 0,3991179 | 0,9999294 | 7,789710969 |
| 8 | 0,6 | INTERMEDIATE | 0,0025 | 200 | 0,0081938 | 0,4184266 | 0,9999173 | 6,1322989 |
| 8 | 0,6 | INTERMEDIATE | 0,0025 | 250 | 0,0123472 | 0,3964816 | 0,9998791 | 5,264401358 |
| 8 | 0,6 | INTERMEDIATE | 0,0025 | 500 | 0,0213709 | 0,3678298 | 0,9999405 | 3,927623103 |
| 8 | 0,6 | INTERMEDIATE | 0,0025 | 1000 | 0,0293416 | 0,3718675 | 0,9998989 | 3,047225361 |
| 8 | 0,6 | INTERMEDIATE | 0,005 | 125 | 0,0032078 | 0,450945 | 0,999833 | 7,99026815 |
| 8 | 0,6 | INTERMEDIATE | 0,005 | 200 | 0,0078418 | 0,4319836 | 0,9998918 | 6,120532128 |
| 8 | 0,6 | INTERMEDIATE | 0,005 | 250 | 0,0099596 | 0,4333499 | 0,9999448 | 5,540256009 |
| 8 | 0,6 | INTERMEDIATE | 0,005 | 500 | 0,0186923 | 0,4426855 | 0,9999221 | 4,004941134 |
| 8 | 0,6 | INTERMEDIATE | 0,005 | 1000 | 0,0272801 | 0,405499 | 0,9999299 | 3,193323952 |
| 8 | 0,6 | INTERMEDIATE | 0,01 | 125 | 0,0022809 | 0,5118935 | 0,999822 | 7,967318772 |
| 8 | 0,6 | INTERMEDIATE | 0,01 | 200 | 0,0062802 | 0,4892113 | 0,9998288 | 6,132154002 |
| 8 | 0,6 | INTERMEDIATE | 0,01 | 250 | 0,0082511 | 0,49684 | 0,9999043 | 5,505862068 |
| 8 | 0,6 | INTERMEDIATE | 0,01 | 500 | 0,014504 | 0,4827749 | 0,9997979 | 4,400284736 |
| 8 | 0,6 | INTERMEDIATE | 0,01 | 1000 | 0,0234627 | 0,4653846 | 0,9998167 | 3,417214326 |
| 8 | 0,6 | INTERMEDIATE | 0,02 | 125 | 0,0018637 | 0,5597234 | 0,9996445 | 7,813416712 |
| 8 | 0,6 | INTERMEDIATE | 0,02 | 200 | 0,0036928 | 0,5958278 | 0,9996662 | 6,282714151 |
| 8 | 0,6 | INTERMEDIATE | 0,02 | 250 | 0,0056268 | 0,58583 | 0,9996147 | 5,627677678 |
| 8 | 0,6 | INTERMEDIATE | 0,02 | 500 | 0,0118846 | 0,5666325 | 0,999648 | 4,403519334 |
| 8 | 0,6 | INTERMEDIATE | 0,02 | 1000 | 0,0182363 | 0,5635001 | 0,9996216 | 3,633592067 |
| 8 | 0,6 | INTERMEDIATE | 0,04 | 125 | 0,0008541 | 0,688818 | 0,9988403 | 7,794573748 |
| 8 | 0,6 | INTERMEDIATE | 0,04 | 200 | 0,0020555 | 0,7206926 | 0,9988751 | 6,285396765 |
| 8 | 0,6 | INTERMEDIATE | 0,04 | 250 | 0,0030336 | 0,7220218 | 0,9988358 | 5,731067023 |
| 8 | 0,6 | INTERMEDIATE | 0,04 | 500 | 0,0072893 | 0,7238257 | 0,9988405 | 4,487927041 |
| 8 | 0,6 | INTERMEDIATE | 0,04 | 1000 | 0,0133955 | 0,7019351 | 0,998646 | 3,693352097 |
| 8 | 0,6 | UNIFORM | 0 | 125 | 0,0054297 | 0,3466774 | 0,9999189 | 8,064447319 |
| 8 | 0,6 | UNIFORM | 0 | 200 | 0,0109374 | 0,3535811 | 0,9997698 | 5,922791223 |
| 8 | 0,6 | UNIFORM | 0 | 250 | 0,01462 | 0,3444248 | 0,9998496 | 5,131753361 |
| 8 | 0,6 | UNIFORM | 0 | 500 | 0,0238964 | 0,3513312 | 0,9999185 | 3,653897899 |
| 8 | 0,6 | UNIFORM | 0 | 1000 | 0,0326928 | 0,330968 | 0,9999608 | 2,744062073 |
| 8 | 0,6 | UNIFORM | 0,0025 | 125 | 0,0051988 | 0,3822089 | 0,9998818 | 7,693672378 |
| 8 | 0,6 | UNIFORM | 0,0025 | 200 | 0,0085832 | 0,4006617 | 0,9999283 | 6,173190002 |
| 8 | 0,6 | UNIFORM | 0,0025 | 250 | 0,0098926 | 0,411033 | 0,9998954 | 5,725318211 |
| 8 | 0,6 | UNIFORM | 0,0025 | 500 | 0,0219405 | 0,3517862 | 0,9999475 | 3,900527561 |
| 8 | 0,6 | UNIFORM | 0,0025 | 1000 | 0,0292092 | 0,3735587 | 0,999898 | 3,057922327 |
| 8 | 0,6 | UNIFORM | 0,005 | 125 | 0,0038186 | 0,4287368 | 0,9999185 | 7,869529332 |
| 8 | 0,6 | UNIFORM | 0,005 | 200 | 0,0066983 | 0,4524249 | 0,9998818 | 6,307268538 |
| 8 | 0,6 | UNIFORM | 0,005 | 250 | 0,0086668 | 0,4690424 | 0,9999397 | 5,598049242 |
| 8 | 0,6 | UNIFORM | 0,005 | 500 | 0,0172464 | 0,4242956 | 0,999906 | 4,271099594 |
| 8 | 0,6 | UNIFORM | 0,005 | 1000 | 0,0271783 | 0,4070492 | 0,9999311 | 3,200072042 |
| 8 | 0,6 | UNIFORM | 0,01 | 125 | 0,0028762 | 0,4734692 | 0,9998807 | 7,94939571 |
| 8 | 0,6 | UNIFORM | 0,01 | 200 | 0,005142 | 0,5082074 | 0,999729 | 6,382578199 |
| 8 | 0,6 | UNIFORM | 0,01 | 250 | 0,0083293 | 0,4731986 | 0,9998849 | 5,654457908 |
| 8 | 0,6 | UNIFORM | 0,01 | 500 | 0,0156049 | 0,4678967 | 0,9998595 | 4,309618468 |
| 8 | 0,6 | UNIFORM | 0,01 | 1000 | 0,0234883 | 0,4653121 | 0,9998145 | 3,414990393 |
| 8 | 0,6 | UNIFORM | 0,02 | 125 | 0,0014808 | 0,5883946 | 0,9995774 | 7,913253961 |
| 8 | 0,6 | UNIFORM | 0,02 | 200 | 0,0038053 | 0,5910848 | 0,9996927 | 6,267702982 |
| 8 | 0,6 | UNIFORM | 0,02 | 250 | 0,0055495 | 0,590609 | 0,999801 | 5,620136278 |
| 8 | 0,6 | UNIFORM | 0,02 | 500 | 0,0115776 | 0,5730537 | 0,9996248 | 4,421505395 |
| 8 | 0,6 | UNIFORM | 0,02 | 1000 | 0,0180748 | 0,5643913 | 0,9996058 | 3,647026302 |
| 8 | 0,6 | UNIFORM | 0,04 | 125 | 0,000837 | 0,712426 | 0,9987156 | 7,612442456 |
| 8 | 0,6 | UNIFORM | 0,04 | 200 | 0,0021 | 0,7137839 | 0,9990489 | 6,30229009 |
| 8 | 0,6 | UNIFORM | 0,04 | 250 | 0,0029459 | 0,721059 | 0,9988476 | 5,778046518 |
| 8 | 0,6 | UNIFORM | 0,04 | 500 | 0,0076885 | 0,7102001 | 0,9991054 | 4,469853159 |
| 8 | 0,6 | UNIFORM | 0,04 | 1000 | 0,0133205 | 0,7046259 | 0,9987168 | 3,693044135 |
| 8 | 0,8 | COMPACT | 0 | 125 | 0,0320406 | 0,6294018 | 0,9998683 | 2,510887679 |
| 8 | 0,8 | COMPACT | 0 | 200 | 0,0373798 | 0,7968023 | 0,9998874 | 2,088895515 |
| 8 | 0,8 | COMPACT | 0 | 250 | 0,0507592 | 0,756528 | 0,9998365 | 1,7152373 |
| 8 | 0,8 | COMPACT | 0 | 500 | 0,0871103 | 0,8470224 | 0,9999373 | 1,025115747 |
| 8 | 0,8 | COMPACT | 0 | 1000 | 0,1511214 | 0,7507841 | 0,9999908 | 0,317191621 |
| 8 | 0,8 | COMPACT | 0,0025 | 125 | 0,0185589 | 0,8808642 | 0,9998518 | 2,827362017 |
| 8 | 0,8 | COMPACT | 0,0025 | 200 | 0,0278922 | 0,9666224 | 0,9999612 | 2,240794142 |
| 8 | 0,8 | COMPACT | 0,0025 | 250 | 0,0350716 | 0,9871739 | 0,9999699 | 1,976528579 |
| 8 | 0,8 | COMPACT | 0,0025 | 500 | 0,0729265 | 0,9865337 | 0,999917 | 1,215513245 |
| 8 | 0,8 | COMPACT | 0,0025 | 1000 | 0,1333565 | 0,8923007 | 0,9998684 | 0,569215113 |
| 8 | 0,8 | COMPACT | 0,005 | 125 | 0,0163032 | 0,9999995 | 0,9997443 | 2,754530993 |
| 8 | 0,8 | COMPACT | 0,005 | 200 | 0,0292995 | 0,9999992 | 0,9994833 | 2,150225499 |
| 8 | 0,8 | COMPACT | 0,005 | 250 | 0,0383235 | 1 | 0,9995508 | 1,87295668 |
| 8 | 0,8 | COMPACT | 0,005 | 500 | 0,0782114 | 0,9999933 | 0,999642 | 1,142060317 |
| 8 | 0,8 | COMPACT | 0,005 | 1000 | 0,1234159 | 0,9999971 | 0,999825 | 0,690071089 |
| 8 | 0,8 | COMPACT | 0,01 | 125 | 0,0207381 | 0,9999999 | 0,9980754 | 2,506840403 |
| 8 | 0,8 | COMPACT | 0,01 | 200 | 0,0330904 | 1 | 0,998357 | 2,024560182 |
| 8 | 0,8 | COMPACT | 0,01 | 250 | 0,0439095 | 0,9999995 | 0,9979274 | 1,732590812 |
| 8 | 0,8 | COMPACT | 0,01 | 500 | 0,0866497 | 0,9999998 | 0,9981924 | 1,038903424 |
| 8 | 0,8 | COMPACT | 0,01 | 1000 | 0,1346071 | 0,9999881 | 0,9991577 | 0,606758882 |
| 8 | 0,8 | COMPACT | 0,02 | 125 | 0,0278443 | 0,9999998 | 0,9942675 | 2,202834352 |
| 8 | 0,8 | COMPACT | 0,02 | 200 | 0,043784 | 1 | 0,9929757 | 1,735539685 |
| 8 | 0,8 | COMPACT | 0,02 | 250 | 0,0539924 | 0,9999998 | 0,9936466 | 1,519890627 |
| 8 | 0,8 | COMPACT | 0,02 | 500 | 0,0977185 | 1 | 0,9948501 | 0,918911869 |
| 8 | 0,8 | COMPACT | 0,02 | 1000 | 0,1504885 | 0,9999801 | 0,9974427 | 0,50143957 |
| 8 | 0,8 | COMPACT | 0,04 | 125 | 0,037842 | 1 | 0,9864427 | 1,886007127 |
| 8 | 0,8 | COMPACT | 0,04 | 200 | 0,0554087 | 1 | 0,9837827 | 1,493317548 |
| 8 | 0,8 | COMPACT | 0,04 | 250 | 0,0659335 | 1 | 0,9855102 | 1,315421077 |
| 8 | 0,8 | COMPACT | 0,04 | 500 | 0,1142856 | 1 | 0,9908363 | 0,76470215 |
| 8 | 0,8 | COMPACT | 0,04 | 1000 | 0,174189 | 0,9999999 | 0,9921515 | 0,366769487 |
| 8 | 0,8 | INTERMEDIATE | 0 | 125 | 0,0320406 | 0,6294018 | 0,9998683 | 2,510887679 |
| 8 | 0,8 | INTERMEDIATE | 0 | 200 | 0,0373798 | 0,7968023 | 0,9998874 | 2,088895515 |
| 8 | 0,8 | INTERMEDIATE | 0 | 250 | 0,0507592 | 0,756528 | 0,9998365 | 1,7152373 |
| 8 | 0,8 | INTERMEDIATE | 0 | 500 | 0,0871103 | 0,8470224 | 0,9999373 | 1,025115747 |
| 8 | 0,8 | INTERMEDIATE | 0 | 1000 | 0,1511214 | 0,7507841 | 0,9999908 | 0,317191621 |
| 8 | 0,8 | INTERMEDIATE | 0,0025 | 125 | 0,0185371 | 0,9002253 | 0,999908 | 2,792794432 |
| 8 | 0,8 | INTERMEDIATE | 0,0025 | 200 | 0,0268745 | 0,986521 | 0,9999477 | 2,255830708 |
| 8 | 0,8 | INTERMEDIATE | 0,0025 | 250 | 0,0355903 | 0,9712643 | 0,999941 | 1,976023562 |
| 8 | 0,8 | INTERMEDIATE | 0,0025 | 500 | 0,0729252 | 0,9865225 | 0,999917 | 1,215533159 |
| 8 | 0,8 | INTERMEDIATE | 0,0025 | 1000 | 0,1333615 | 0,8924914 | 0,9998689 | 0,569277999 |
| 8 | 0,8 | INTERMEDIATE | 0,005 | 125 | 0,0157121 | 0,9999995 | 0,9997815 | 2,792494065 |
| 8 | 0,8 | INTERMEDIATE | 0,005 | 200 | 0,0299504 | 1 | 0,9996272 | 2,12753325 |
| 8 | 0,8 | INTERMEDIATE | 0,005 | 250 | 0,0381003 | 1 | 0,9995061 | 1,878986611 |
| 8 | 0,8 | INTERMEDIATE | 0,005 | 500 | 0,077815 | 0,9999937 | 0,9997588 | 1,147195247 |
| 8 | 0,8 | INTERMEDIATE | 0,005 | 1000 | 0,1233293 | 0,9999999 | 0,9998247 | 0,69074933 |
| 8 | 0,8 | INTERMEDIATE | 0,01 | 125 | 0,0206611 | 0,9999999 | 0,9980084 | 2,510674045 |
| 8 | 0,8 | INTERMEDIATE | 0,01 | 200 | 0,034265 | 1 | 0,9976754 | 1,988536253 |
| 8 | 0,8 | INTERMEDIATE | 0,01 | 250 | 0,0431881 | 0,9999997 | 0,9980118 | 1,749667516 |
| 8 | 0,8 | INTERMEDIATE | 0,01 | 500 | 0,086069 | 0,9999998 | 0,9982341 | 1,045650237 |
| 8 | 0,8 | INTERMEDIATE | 0,01 | 1000 | 0,1346728 | 0,999988 | 0,9991524 | 0,606293642 |
| 8 | 0,8 | INTERMEDIATE | 0,02 | 125 | 0,0282438 | 0,9999998 | 0,9939472 | 2,188121782 |
| 8 | 0,8 | INTERMEDIATE | 0,02 | 200 | 0,0442478 | 0,9999998 | 0,9947994 | 1,724679123 |
| 8 | 0,8 | INTERMEDIATE | 0,02 | 250 | 0,054081 | 0,9999998 | 0,9935568 | 1,518208082 |
| 8 | 0,8 | INTERMEDIATE | 0,02 | 500 | 0,0978036 | 1 | 0,9947962 | 0,918047857 |
| 8 | 0,8 | INTERMEDIATE | 0,02 | 1000 | 0,1502211 | 0,999987 | 0,997493 | 0,503105502 |
| 8 | 0,8 | INTERMEDIATE | 0,04 | 125 | 0,0386105 | 1 | 0,9864642 | 1,865253944 |
| 8 | 0,8 | INTERMEDIATE | 0,04 | 200 | 0,0556189 | 1 | 0,9837185 | 1,489434467 |
| 8 | 0,8 | INTERMEDIATE | 0,04 | 250 | 0,065047 | 1 | 0,9854465 | 1,329225092 |
| 8 | 0,8 | INTERMEDIATE | 0,04 | 500 | 0,1149821 | 1 | 0,9860424 | 0,758774933 |
| 8 | 0,8 | INTERMEDIATE | 0,04 | 1000 | 0,173815 | 0,9999999 | 0,9921996 | 0,368717418 |
| 8 | 0,8 | UNIFORM | 0 | 125 | 0,0320406 | 0,6294018 | 0,9998683 | 2,510887679 |
| 8 | 0,8 | UNIFORM | 0 | 200 | 0,0373798 | 0,7968023 | 0,9998874 | 2,088895515 |
| 8 | 0,8 | UNIFORM | 0 | 250 | 0,0507592 | 0,756528 | 0,9998365 | 1,7152373 |
| 8 | 0,8 | UNIFORM | 0 | 500 | 0,0871103 | 0,8470224 | 0,9999373 | 1,025115747 |
| 8 | 0,8 | UNIFORM | 0 | 1000 | 0,1511214 | 0,7507841 | 0,9999908 | 0,317191621 |
| 8 | 0,8 | UNIFORM | 0,0025 | 125 | 0,0185266 | 0,882247 | 0,9998687 | 2,826801325 |
| 8 | 0,8 | UNIFORM | 0,0025 | 200 | 0,0273418 | 0,9807001 | 0,9999617 | 2,244839614 |
| 8 | 0,8 | UNIFORM | 0,0025 | 250 | 0,0345037 | 0,9979503 | 0,9999591 | 1,983313299 |
| 8 | 0,8 | UNIFORM | 0,0025 | 500 | 0,0728917 | 0,9869644 | 0,9999172 | 1,215920934 |
| 8 | 0,8 | UNIFORM | 0,0025 | 1000 | 0,133362 | 0,8922649 | 0,9998686 | 0,569152134 |
| 8 | 0,8 | UNIFORM | 0,005 | 125 | 0,0157106 | 1 | 0,9997101 | 2,792587139 |
| 8 | 0,8 | UNIFORM | 0,005 | 200 | 0,0296358 | 0,9999997 | 0,9993957 | 2,138440013 |
| 8 | 0,8 | UNIFORM | 0,005 | 250 | 0,0383369 | 1 | 0,999535 | 1,87259459 |
| 8 | 0,8 | UNIFORM | 0,005 | 500 | 0,0781502 | 0,9999935 | 0,9996429 | 1,142851829 |
| 8 | 0,8 | UNIFORM | 0,005 | 1000 | 0,1233198 | 0,9999997 | 0,9998262 | 0,690824343 |
| 8 | 0,8 | UNIFORM | 0,01 | 125 | 0,0207707 | 0,9999999 | 0,9979976 | 2,505216955 |
| 8 | 0,8 | UNIFORM | 0,01 | 200 | 0,0333057 | 1 | 0,99829 | 2,017861964 |
| 8 | 0,8 | UNIFORM | 0,01 | 250 | 0,0439568 | 0,9999994 | 0,9978314 | 1,731479238 |
| 8 | 0,8 | UNIFORM | 0,01 | 500 | 0,0866509 | 0,9999998 | 0,998213 | 1,038889371 |
| 8 | 0,8 | UNIFORM | 0,01 | 1000 | 0,1345762 | 0,9999882 | 0,9991583 | 0,606977559 |
| 8 | 0,8 | UNIFORM | 0,02 | 125 | 0,028289 | 0,9999998 | 0,9938833 | 2,18647268 |
| 8 | 0,8 | UNIFORM | 0,02 | 200 | 0,0440632 | 1 | 0,9931518 | 1,728986954 |
| 8 | 0,8 | UNIFORM | 0,02 | 250 | 0,0542992 | 0,9999998 | 0,9935557 | 1,514074436 |
| 8 | 0,8 | UNIFORM | 0,02 | 500 | 0,0976845 | 1 | 0,9947647 | 0,919257636 |
| 8 | 0,8 | UNIFORM | 0,02 | 1000 | 0,1505033 | 0,9999819 | 0,9974609 | 0,501349019 |
| 8 | 0,8 | UNIFORM | 0,04 | 125 | 0,0385657 | 0,9999975 | 0,9801741 | 1,866455496 |
| 8 | 0,8 | UNIFORM | 0,04 | 200 | 0,0553406 | 1 | 0,988515 | 1,494578726 |
| 8 | 0,8 | UNIFORM | 0,04 | 250 | 0,0655054 | 0,9999991 | 0,9852108 | 1,322063204 |
| 8 | 0,8 | UNIFORM | 0,04 | 500 | 0,115522 | 1 | 0,9858367 | 0,75420853 |
| 8 | 0,8 | UNIFORM | 0,04 | 1000 | 0,173785 | 0,9999999 | 0,9921881 | 0,368873671 |
| 4 | 0,6 | COMPACT | 0 | 125 | 0,0204385 | 0,5381895 | 0,9998883 | 3,497558603 |
| 4 | 0,6 | COMPACT | 0 | 200 | 0,0279568 | 0,5997619 | 0,9999282 | 2,786142244 |
| 4 | 0,6 | COMPACT | 0 | 250 | 0,0363631 | 0,5847612 | 0,9998348 | 2,351689597 |
| 4 | 0,6 | COMPACT | 0 | 500 | 0,0649346 | 0,5565343 | 0,9999557 | 1,3377824 |
| 4 | 0,6 | COMPACT | 0 | 1000 | 0,092351 | 0,5533892 | 0,9999947 | 0,73327746 |
| 4 | 0,6 | COMPACT | 0,0025 | 125 | 0,0174676 | 0,5914379 | 0,9999864 | 3,621667136 |
| 4 | 0,6 | COMPACT | 0,0025 | 200 | 0,0244055 | 0,6588797 | 0,9999805 | 2,895973666 |
| 4 | 0,6 | COMPACT | 0,0025 | 250 | 0,0309164 | 0,6541563 | 0,9999652 | 2,532726557 |
| 4 | 0,6 | COMPACT | 0,0025 | 500 | 0,054816 | 0,6698565 | 0,9999708 | 1,636772981 |
| 4 | 0,6 | COMPACT | 0,0025 | 1000 | 0,0848862 | 0,6150012 | 0,9999602 | 0,947802008 |
| 4 | 0,6 | COMPACT | 0,005 | 125 | 0,0143723 | 0,6556659 | 0,9999281 | 3,736191982 |
| 4 | 0,6 | COMPACT | 0,005 | 200 | 0,0224672 | 0,6911382 | 0,9999859 | 2,955868592 |
| 4 | 0,6 | COMPACT | 0,005 | 250 | 0,0282059 | 0,6987984 | 0,9999798 | 2,603661519 |
| 4 | 0,6 | COMPACT | 0,005 | 500 | 0,052773 | 0,6897248 | 0,9999633 | 1,687998902 |
| 4 | 0,6 | COMPACT | 0,005 | 1000 | 0,080842 | 0,6503642 | 0,999927 | 1,050425885 |
| 4 | 0,6 | COMPACT | 0,01 | 125 | 0,0123368 | 0,717938 | 0,9997953 | 3,76138246 |
| 4 | 0,6 | COMPACT | 0,01 | 200 | 0,0189061 | 0,7675009 | 0,9999176 | 3,034936559 |
| 4 | 0,6 | COMPACT | 0,01 | 250 | 0,0234247 | 0,7719883 | 0,9999076 | 2,738474897 |
| 4 | 0,6 | COMPACT | 0,01 | 500 | 0,0471853 | 0,754857 | 0,9999122 | 1,814725908 |
| 4 | 0,6 | COMPACT | 0,01 | 1000 | 0,0741549 | 0,7113478 | 0,9998608 | 1,20183768 |
| 4 | 0,6 | COMPACT | 0,02 | 125 | 0,008607 | 0,8565565 | 0,99957 | 3,79510335 |
| 4 | 0,6 | COMPACT | 0,02 | 200 | 0,0142306 | 0,8953549 | 0,9997736 | 3,105663645 |
| 4 | 0,6 | COMPACT | 0,02 | 250 | 0,0185539 | 0,8995388 | 0,9998327 | 2,793011325 |
| 4 | 0,6 | COMPACT | 0,02 | 500 | 0,03901 | 0,8747611 | 0,9997576 | 1,962388753 |
| 4 | 0,6 | COMPACT | 0,02 | 1000 | 0,0643318 | 0,8187541 | 0,9995697 | 1,389188153 |
| 4 | 0,6 | COMPACT | 0,04 | 125 | 0,0063442 | 0,9963596 | 0,9992769 | 3,729526412 |
| 4 | 0,6 | COMPACT | 0,04 | 200 | 0,0120151 | 0,9999997 | 0,9994718 | 3,067775055 |
| 4 | 0,6 | COMPACT | 0,04 | 250 | 0,0158148 | 1 | 0,9994435 | 2,785793732 |
| 4 | 0,6 | COMPACT | 0,04 | 500 | 0,0315805 | 0,9958337 | 0,9995996 | 2,077140359 |
| 4 | 0,6 | COMPACT | 0,04 | 1000 | 0,0528159 | 0,9651684 | 0,9995307 | 1,560468259 |
| 4 | 0,6 | INTERMEDIATE | 0 | 125 | 0,0204385 | 0,5381895 | 0,9998883 | 3,497558603 |
| 4 | 0,6 | INTERMEDIATE | 0 | 200 | 0,0279568 | 0,5997619 | 0,9999282 | 2,786142244 |
| 4 | 0,6 | INTERMEDIATE | 0 | 250 | 0,0363631 | 0,5847612 | 0,9998348 | 2,351689597 |
| 4 | 0,6 | INTERMEDIATE | 0 | 500 | 0,0649346 | 0,5565343 | 0,9999557 | 1,3377824 |
| 4 | 0,6 | INTERMEDIATE | 0 | 1000 | 0,092351 | 0,5533892 | 0,9999947 | 0,73327746 |
| 4 | 0,6 | INTERMEDIATE | 0,0025 | 125 | 0,0155527 | 0,6254245 | 0,9999846 | 3,70877719 |
| 4 | 0,6 | INTERMEDIATE | 0,0025 | 200 | 0,0240901 | 0,6589081 | 0,9999838 | 2,916294764 |
| 4 | 0,6 | INTERMEDIATE | 0,0025 | 250 | 0,0312797 | 0,6531215 | 0,9999604 | 2,515830543 |
| 4 | 0,6 | INTERMEDIATE | 0,0025 | 500 | 0,0549573 | 0,6728879 | 0,9999652 | 1,632303284 |
| 4 | 0,6 | INTERMEDIATE | 0,0025 | 1000 | 0,0849969 | 0,6129622 | 0,9999537 | 0,943796049 |
| 4 | 0,6 | INTERMEDIATE | 0,005 | 125 | 0,0135416 | 0,6761108 | 0,9999105 | 3,760783256 |
| 4 | 0,6 | INTERMEDIATE | 0,005 | 200 | 0,0235961 | 0,6754407 | 0,9999835 | 2,914478765 |
| 4 | 0,6 | INTERMEDIATE | 0,005 | 250 | 0,0276486 | 0,7081278 | 0,9999665 | 2,617767021 |
| 4 | 0,6 | INTERMEDIATE | 0,005 | 500 | 0,0520603 | 0,6929329 | 0,9999429 | 1,706824468 |
| 4 | 0,6 | INTERMEDIATE | 0,005 | 1000 | 0,0818272 | 0,6403371 | 0,9999518 | 1,025288848 |
| 4 | 0,6 | INTERMEDIATE | 0,01 | 125 | 0,0118863 | 0,7310459 | 0,9997846 | 3,771715748 |
| 4 | 0,6 | INTERMEDIATE | 0,01 | 200 | 0,0180375 | 0,7883289 | 0,9999001 | 3,051348183 |
| 4 | 0,6 | INTERMEDIATE | 0,01 | 250 | 0,0246606 | 0,7497037 | 0,9999132 | 2,708692602 |
| 4 | 0,6 | INTERMEDIATE | 0,01 | 500 | 0,047032 | 0,757085 | 0,9999045 | 1,817757088 |
| 4 | 0,6 | INTERMEDIATE | 0,01 | 1000 | 0,0733226 | 0,7180513 | 0,9998552 | 1,219038121 |
| 4 | 0,6 | INTERMEDIATE | 0,02 | 125 | 0,0088588 | 0,8541469 | 0,9995756 | 3,767911259 |
| 4 | 0,6 | INTERMEDIATE | 0,02 | 200 | 0,0143877 | 0,8891388 | 0,9997324 | 3,106627317 |
| 4 | 0,6 | INTERMEDIATE | 0,02 | 250 | 0,0184379 | 0,90629 | 0,9997694 | 2,787837615 |
| 4 | 0,6 | INTERMEDIATE | 0,02 | 500 | 0,0398028 | 0,8601473 | 0,9997647 | 1,951450455 |
| 4 | 0,6 | INTERMEDIATE | 0,02 | 1000 | 0,0642513 | 0,8172457 | 0,9995541 | 1,391020398 |
| 4 | 0,6 | INTERMEDIATE | 0,04 | 125 | 0,0064386 | 0,9967933 | 0,9993341 | 3,713269719 |
| 4 | 0,6 | INTERMEDIATE | 0,04 | 200 | 0,0120501 | 1 | 0,9994801 | 3,064794278 |
| 4 | 0,6 | INTERMEDIATE | 0,04 | 250 | 0,0158958 | 1 | 0,9995158 | 2,78054443 |
| 4 | 0,6 | INTERMEDIATE | 0,04 | 500 | 0,0319395 | 0,9999803 | 0,9995932 | 2,061142129 |
| 4 | 0,6 | INTERMEDIATE | 0,04 | 1000 | 0,0516396 | 0,9789383 | 0,9994968 | 1,577001094 |
| 4 | 0,6 | UNIFORM | 0 | 125 | 0,0204385 | 0,5381895 | 0,9998883 | 3,497558603 |
| 4 | 0,6 | UNIFORM | 0 | 200 | 0,0279568 | 0,5997619 | 0,9999282 | 2,786142244 |
| 4 | 0,6 | UNIFORM | 0 | 250 | 0,0363631 | 0,5847612 | 0,9998348 | 2,351689597 |
| 4 | 0,6 | UNIFORM | 0 | 500 | 0,0649346 | 0,5565343 | 0,9999557 | 1,3377824 |
| 4 | 0,6 | UNIFORM | 0 | 1000 | 0,092351 | 0,5533892 | 0,9999947 | 0,73327746 |
| 4 | 0,6 | UNIFORM | 0,0025 | 125 | 0,0160867 | 0,6116422 | 0,9999884 | 3,697773848 |
| 4 | 0,6 | UNIFORM | 0,0025 | 200 | 0,0238033 | 0,6764373 | 0,9999843 | 2,899085219 |
| 4 | 0,6 | UNIFORM | 0,0025 | 250 | 0,0305229 | 0,6644968 | 0,9999672 | 2,537474009 |
| 4 | 0,6 | UNIFORM | 0,0025 | 500 | 0,0553181 | 0,6685989 | 0,9999625 | 1,623287354 |
| 4 | 0,6 | UNIFORM | 0,0025 | 1000 | 0,0849033 | 0,6113494 | 0,9999546 | 0,943907644 |
| 4 | 0,6 | UNIFORM | 0,005 | 125 | 0,0138754 | 0,6679088 | 0,999919 | 3,750565368 |
| 4 | 0,6 | UNIFORM | 0,005 | 200 | 0,0227269 | 0,6885896 | 0,9999813 | 2,944031234 |
| 4 | 0,6 | UNIFORM | 0,005 | 250 | 0,030034 | 0,6644741 | 0,9999685 | 2,562542388 |
| 4 | 0,6 | UNIFORM | 0,005 | 500 | 0,0536962 | 0,6817065 | 0,9999681 | 1,664800907 |
| 4 | 0,6 | UNIFORM | 0,005 | 1000 | 0,0814401 | 0,6435808 | 0,9999436 | 1,034759323 |
| 4 | 0,6 | UNIFORM | 0,01 | 125 | 0,011484 | 0,7483949 | 0,9998389 | 3,763783086 |
| 4 | 0,6 | UNIFORM | 0,01 | 200 | 0,0188177 | 0,7724327 | 0,9999299 | 3,030382926 |
| 4 | 0,6 | UNIFORM | 0,01 | 250 | 0,0231683 | 0,7784317 | 0,9999093 | 2,741442507 |
| 4 | 0,6 | UNIFORM | 0,01 | 500 | 0,0466769 | 0,7596809 | 0,9999198 | 1,826348764 |
| 4 | 0,6 | UNIFORM | 0,01 | 1000 | 0,0731036 | 0,722495 | 0,9998492 | 1,224078658 |
| 4 | 0,6 | UNIFORM | 0,02 | 125 | 0,0092099 | 0,8406961 | 0,999664 | 3,761586216 |
| 4 | 0,6 | UNIFORM | 0,02 | 200 | 0,0140092 | 0,9064391 | 0,9997823 | 3,099419462 |
| 4 | 0,6 | UNIFORM | 0,02 | 250 | 0,018629 | 0,9008742 | 0,9998095 | 2,785949424 |
| 4 | 0,6 | UNIFORM | 0,02 | 500 | 0,0398539 | 0,8607764 | 0,9997912 | 1,949364678 |
| 4 | 0,6 | UNIFORM | 0,02 | 1000 | 0,0647882 | 0,8134175 | 0,9996734 | 1,381411328 |
| 4 | 0,6 | UNIFORM | 0,04 | 125 | 0,0062198 | 0,9977759 | 0,9994703 | 3,745863179 |
| 4 | 0,6 | UNIFORM | 0,04 | 200 | 0,0118074 | 1 | 0,999503 | 3,085641003 |
| 4 | 0,6 | UNIFORM | 0,04 | 250 | 0,015869 | 1 | 0,9994702 | 2,782277007 |
| 4 | 0,6 | UNIFORM | 0,04 | 500 | 0,0317395 | 0,9902799 | 0,9995909 | 2,077719058 |
| 4 | 0,6 | UNIFORM | 0,04 | 1000 | 0,052274 | 0,9704129 | 0,9994986 | 1,5686753 |
| 4 | 0,8 | COMPACT | 0 | 125 | 0,0204385 | 0,5381895 | 0,9998883 | 3,497558603 |
| 4 | 0,8 | COMPACT | 0 | 200 | 0,0279568 | 0,5997619 | 0,9999282 | 2,786142244 |
| 4 | 0,8 | COMPACT | 0 | 250 | 0,0363631 | 0,5847612 | 0,9998348 | 2,351689597 |
| 4 | 0,8 | COMPACT | 0 | 500 | 0,0649346 | 0,5565343 | 0,9999557 | 1,3377824 |
| 4 | 0,8 | COMPACT | 0 | 1000 | 0,092351 | 0,5533892 | 0,9999947 | 0,73327746 |
| 4 | 0,8 | COMPACT | 0,0025 | 125 | 0,0167916 | 0,6129896 | 0,9999877 | 3,621113459 |
| 4 | 0,8 | COMPACT | 0,0025 | 200 | 0,023227 | 0,6934594 | 0,9999843 | 2,901436131 |
| 4 | 0,8 | COMPACT | 0,0025 | 250 | 0,0295305 | 0,6851325 | 0,9999772 | 2,556770029 |
| 4 | 0,8 | COMPACT | 0,0025 | 500 | 0,0569044 | 0,6504653 | 0,9999547 | 1,582421441 |
| 4 | 0,8 | COMPACT | 0,0025 | 1000 | 0,0841676 | 0,6306902 | 0,9999506 | 0,975085412 |
| 4 | 0,8 | COMPACT | 0,005 | 125 | 0,0137828 | 0,6812613 | 0,9999554 | 3,717128242 |
| 4 | 0,8 | COMPACT | 0,005 | 200 | 0,0205902 | 0,7398442 | 0,9999727 | 2,978123517 |
| 4 | 0,8 | COMPACT | 0,005 | 250 | 0,0269212 | 0,7295687 | 0,9999828 | 2,620765438 |
| 4 | 0,8 | COMPACT | 0,005 | 500 | 0,0527457 | 0,6954401 | 0,9999489 | 1,686887254 |
| 4 | 0,8 | COMPACT | 0,005 | 1000 | 0,0805615 | 0,667378 | 0,9999297 | 1,066098573 |
| 4 | 0,8 | COMPACT | 0,01 | 125 | 0,0109554 | 0,7879617 | 0,9998428 | 3,703333855 |
| 4 | 0,8 | COMPACT | 0,01 | 200 | 0,0169662 | 0,835456 | 0,9998603 | 3,026428335 |
| 4 | 0,8 | COMPACT | 0,01 | 250 | 0,0231195 | 0,8186985 | 0,9999466 | 2,672884405 |
| 4 | 0,8 | COMPACT | 0,01 | 500 | 0,0462397 | 0,7918913 | 0,9999006 | 1,817887608 |
| 4 | 0,8 | COMPACT | 0,01 | 1000 | 0,0722948 | 0,7619647 | 0,9998204 | 1,244445661 |
| 4 | 0,8 | COMPACT | 0,02 | 125 | 0,008295 | 0,9265898 | 0,9996824 | 3,635630523 |
| 4 | 0,8 | COMPACT | 0,02 | 200 | 0,0134948 | 0,9697136 | 0,9997571 | 3,008254151 |
| 4 | 0,8 | COMPACT | 0,02 | 250 | 0,0177266 | 0,9653353 | 0,9997675 | 2,726634322 |
| 4 | 0,8 | COMPACT | 0,02 | 500 | 0,0377478 | 0,9446787 | 0,9997648 | 1,936998438 |
| 4 | 0,8 | COMPACT | 0,02 | 1000 | 0,0637844 | 0,8878976 | 0,9995901 | 1,383225169 |
| 4 | 0,8 | COMPACT | 0,04 | 125 | 0,0092474 | 1 | 0,9988859 | 3,335609825 |
| 4 | 0,8 | COMPACT | 0,04 | 200 | 0,0166675 | 0,9999996 | 0,9988392 | 2,731806156 |
| 4 | 0,8 | COMPACT | 0,04 | 250 | 0,0213977 | 0,9999999 | 0,999147 | 2,474565279 |
| 4 | 0,8 | COMPACT | 0,04 | 500 | 0,0411687 | 1 | 0,9991725 | 1,799054133 |
| 4 | 0,8 | COMPACT | 0,04 | 1000 | 0,0630988 | 1 | 0,9993733 | 1,360265104 |
| 4 | 0,8 | INTERMEDIATE | 0 | 125 | 0,0204385 | 0,5381895 | 0,9998883 | 3,497558603 |
| 4 | 0,8 | INTERMEDIATE | 0 | 200 | 0,0279568 | 0,5997619 | 0,9999282 | 2,786142244 |
| 4 | 0,8 | INTERMEDIATE | 0 | 250 | 0,0363631 | 0,5847612 | 0,9998348 | 2,351689597 |
| 4 | 0,8 | INTERMEDIATE | 0 | 500 | 0,0649346 | 0,5565343 | 0,9999557 | 1,3377824 |
| 4 | 0,8 | INTERMEDIATE | 0 | 1000 | 0,092351 | 0,5533892 | 0,9999947 | 0,73327746 |
| 4 | 0,8 | INTERMEDIATE | 0,0025 | 125 | 0,0167916 | 0,6129896 | 0,9999877 | 3,621113459 |
| 4 | 0,8 | INTERMEDIATE | 0,0025 | 200 | 0,023227 | 0,6934594 | 0,9999843 | 2,901436131 |
| 4 | 0,8 | INTERMEDIATE | 0,0025 | 250 | 0,0295305 | 0,6851325 | 0,9999772 | 2,556770029 |
| 4 | 0,8 | INTERMEDIATE | 0,0025 | 500 | 0,0569044 | 0,6504653 | 0,9999547 | 1,582421441 |
| 4 | 0,8 | INTERMEDIATE | 0,0025 | 1000 | 0,0841676 | 0,6306902 | 0,9999506 | 0,975085412 |
| 4 | 0,8 | INTERMEDIATE | 0,005 | 125 | 0,0137828 | 0,6812613 | 0,9999554 | 3,717128242 |
| 4 | 0,8 | INTERMEDIATE | 0,005 | 200 | 0,0205902 | 0,7398442 | 0,9999727 | 2,978123517 |
| 4 | 0,8 | INTERMEDIATE | 0,005 | 250 | 0,0269212 | 0,7295687 | 0,9999828 | 2,620765438 |
| 4 | 0,8 | INTERMEDIATE | 0,005 | 500 | 0,0527457 | 0,6954401 | 0,9999489 | 1,686887254 |
| 4 | 0,8 | INTERMEDIATE | 0,005 | 1000 | 0,0805615 | 0,667378 | 0,9999297 | 1,066098573 |
| 4 | 0,8 | INTERMEDIATE | 0,01 | 125 | 0,0109554 | 0,7879617 | 0,9998428 | 3,703333855 |
| 4 | 0,8 | INTERMEDIATE | 0,01 | 200 | 0,0169662 | 0,835456 | 0,9998603 | 3,026428335 |
| 4 | 0,8 | INTERMEDIATE | 0,01 | 250 | 0,0231195 | 0,8186985 | 0,9999466 | 2,672884405 |
| 4 | 0,8 | INTERMEDIATE | 0,01 | 500 | 0,0462397 | 0,7918913 | 0,9999006 | 1,817887608 |
| 4 | 0,8 | INTERMEDIATE | 0,01 | 1000 | 0,0722948 | 0,7619647 | 0,9998204 | 1,244445661 |
| 4 | 0,8 | INTERMEDIATE | 0,02 | 125 | 0,008295 | 0,9265898 | 0,9996824 | 3,635630523 |
| 4 | 0,8 | INTERMEDIATE | 0,02 | 200 | 0,0134948 | 0,9697136 | 0,9997571 | 3,008254151 |
| 4 | 0,8 | INTERMEDIATE | 0,02 | 250 | 0,0177266 | 0,9653353 | 0,9997675 | 2,726634322 |
| 4 | 0,8 | INTERMEDIATE | 0,02 | 500 | 0,0377478 | 0,9446787 | 0,9997648 | 1,936998438 |
| 4 | 0,8 | INTERMEDIATE | 0,02 | 1000 | 0,0637844 | 0,8878976 | 0,9995901 | 1,383225169 |
| 4 | 0,8 | INTERMEDIATE | 0,04 | 125 | 0,0092474 | 1 | 0,9988859 | 3,335609825 |
| 4 | 0,8 | INTERMEDIATE | 0,04 | 200 | 0,0166675 | 0,9999996 | 0,9988392 | 2,731806156 |
| 4 | 0,8 | INTERMEDIATE | 0,04 | 250 | 0,0213977 | 0,9999999 | 0,999147 | 2,474565279 |
| 4 | 0,8 | INTERMEDIATE | 0,04 | 500 | 0,0411687 | 1 | 0,9991725 | 1,799054133 |
| 4 | 0,8 | INTERMEDIATE | 0,04 | 1000 | 0,0630988 | 1 | 0,9993733 | 1,360265104 |
| 4 | 0,8 | UNIFORM | 0 | 125 | 0,0204385 | 0,5381895 | 0,9998883 | 3,497558603 |
| 4 | 0,8 | UNIFORM | 0 | 200 | 0,0279568 | 0,5997619 | 0,9999282 | 2,786142244 |
| 4 | 0,8 | UNIFORM | 0 | 250 | 0,0363631 | 0,5847612 | 0,9998348 | 2,351689597 |
| 4 | 0,8 | UNIFORM | 0 | 500 | 0,0649346 | 0,5565343 | 0,9999557 | 1,3377824 |
| 4 | 0,8 | UNIFORM | 0 | 1000 | 0,092351 | 0,5533892 | 0,9999947 | 0,73327746 |
| 4 | 0,8 | UNIFORM | 0,0025 | 125 | 0,0167916 | 0,6129896 | 0,9999877 | 3,621113459 |
| 4 | 0,8 | UNIFORM | 0,0025 | 200 | 0,023227 | 0,6934594 | 0,9999843 | 2,901436131 |
| 4 | 0,8 | UNIFORM | 0,0025 | 250 | 0,0295305 | 0,6851325 | 0,9999772 | 2,556770029 |
| 4 | 0,8 | UNIFORM | 0,0025 | 500 | 0,0569044 | 0,6504653 | 0,9999547 | 1,582421441 |
| 4 | 0,8 | UNIFORM | 0,0025 | 1000 | 0,0841676 | 0,6306902 | 0,9999506 | 0,975085412 |
| 4 | 0,8 | UNIFORM | 0,005 | 125 | 0,0137828 | 0,6812613 | 0,9999554 | 3,717128242 |
| 4 | 0,8 | UNIFORM | 0,005 | 200 | 0,0205902 | 0,7398442 | 0,9999727 | 2,978123517 |
| 4 | 0,8 | UNIFORM | 0,005 | 250 | 0,0269212 | 0,7295687 | 0,9999828 | 2,620765438 |
| 4 | 0,8 | UNIFORM | 0,005 | 500 | 0,0527457 | 0,6954401 | 0,9999489 | 1,686887254 |
| 4 | 0,8 | UNIFORM | 0,005 | 1000 | 0,0805615 | 0,667378 | 0,9999297 | 1,066098573 |
| 4 | 0,8 | UNIFORM | 0,01 | 125 | 0,0109554 | 0,7879617 | 0,9998428 | 3,703333855 |
| 4 | 0,8 | UNIFORM | 0,01 | 200 | 0,0169662 | 0,835456 | 0,9998603 | 3,026428335 |
| 4 | 0,8 | UNIFORM | 0,01 | 250 | 0,0231195 | 0,8186985 | 0,9999466 | 2,672884405 |
| 4 | 0,8 | UNIFORM | 0,01 | 500 | 0,0462397 | 0,7918913 | 0,9999006 | 1,817887608 |
| 4 | 0,8 | UNIFORM | 0,01 | 1000 | 0,0722948 | 0,7619647 | 0,9998204 | 1,244445661 |
| 4 | 0,8 | UNIFORM | 0,02 | 125 | 0,008295 | 0,9265898 | 0,9996824 | 3,635630523 |
| 4 | 0,8 | UNIFORM | 0,02 | 200 | 0,0134948 | 0,9697136 | 0,9997571 | 3,008254151 |
| 4 | 0,8 | UNIFORM | 0,02 | 250 | 0,0177266 | 0,9653353 | 0,9997675 | 2,726634322 |
| 4 | 0,8 | UNIFORM | 0,02 | 500 | 0,0377478 | 0,9446787 | 0,9997648 | 1,936998438 |
| 4 | 0,8 | UNIFORM | 0,02 | 1000 | 0,0637844 | 0,8878976 | 0,9995901 | 1,383225169 |
| 4 | 0,8 | UNIFORM | 0,04 | 125 | 0,0092474 | 1 | 0,9988859 | 3,335609825 |
| 4 | 0,8 | UNIFORM | 0,04 | 200 | 0,0166675 | 0,9999996 | 0,9988392 | 2,731806156 |
| 4 | 0,8 | UNIFORM | 0,04 | 250 | 0,0213977 | 0,9999999 | 0,999147 | 2,474565279 |
| 4 | 0,8 | UNIFORM | 0,04 | 500 | 0,0411687 | 1 | 0,9991725 | 1,799054133 |
| 4 | 0,8 | UNIFORM | 0,04 | 1000 | 0,0630988 | 1 | 0,9993733 | 1,360265104 |

*Table 1: Micro-parameter combinations with corresponding induced values of (p, q) and goodness of fit estimates ($r^2$). Take-off time is also calculated for each combination using induced values of (p, q).*